\documentclass{article}

\pdfoutput=1
\usepackage{arxiv}

\usepackage[utf8]{inputenc} % allow utf-8 input
\usepackage[T1]{fontenc}    % use 8-bit T1 fonts
\usepackage{hyperref}       % hyperlinks
\usepackage{url}            % simple URL typesetting
\usepackage{booktabs}       % professional-quality tables
\usepackage{amsfonts}       % blackboard math symbols
\usepackage{nicefrac}       % compact symbols for 1/2, etc.
\usepackage{microtype}      % microtypography
\usepackage{lipsum}
\usepackage{graphicx}
\graphicspath{ {./images/} }
\usepackage{hyperref}
\usepackage{amsmath}
\usepackage{bm}
\usepackage{siunitx}
 \DeclareSIUnit\pN{\pico\newton}
 \DeclareSIUnit[per-mode=symbol]\D{\um\squared\per\second}
 \DeclareSIUnit[inter-unit-product=\ensuremath{{}\cdot{}}]\pNnm{\pico\newton\nano\meter}

\newcommand*{\kT}{k_{\textup{B}}\textup{T}}

\title{Optimal control of nonequilibrium systems through automatic differentiation}

\author{
Megan C. Engel \\
 School of Engineering and Applied Sciences\\
 Harvard University\\ 
 Cambridge, MA 02138 \\
  \texttt{mcengel@seas.harvard.edu} \\
   \And
Jamie A. Smith \\
Google Research\\
Mountain View CA 94043\\
  \texttt{jamieas@google.com} \\
  \And
 Michael P. Brenner \\
 School of Engineering and Applied Sciences\\
 Harvard University\\ 
 Cambridge, MA 02138 \\
 Google Research\\
Mountain View CA 94043\\
  \texttt{brenner@seas.harvard.edu} \\
}

\begin{document}
\maketitle
\begin{abstract}
Controlling the evolution of nonequilibrium systems to minimize dissipated heat or work is a key goal for designing nanodevices, both in nanotechnology and biology. Progress in computing optimal protocols has thus far been limited to either simple systems or near-equilibrium evolution. Here, we present an approach for computing optimal protocols based on automatic differentiation. Our methodology is applicable to complex systems and multidimensional protocols and is valid arbitrarily far from equilibrium. We validate our method by reproducing theoretical optimal protocols for a Brownian particle in a time-varying harmonic trap. We also compute departures from near-equilibrium behaviour for magnetization reversal on an Ising lattice and for barrier crossing driven by a harmonic trap, which has been used to represent a range of biological processes including biomolecular unfolding reactions. Algorithms based on automatic differentiation outperform the near-equilibrium theory for far-from-equilibrium magnetization reversal and driven barrier crossing. The optimal protocol for crossing an energy landscape barrier of 10$\kT$ is found to hasten the approach to, and slow the departure from, the barrier region compared to the near-equilibrium theoretical protocol.
\end{abstract}

% keywords can be removed

\section{Introduction}
The control of nonequilibrium phenomena at microscopic scales is central to biology and nanotechnology.  Evolution has exquisitely tuned cellular processes to perform out-of-equilibrium tasks, ranging from machines like ATP synthase to metabolic factories converting raw materials and energy into functional macro-molecules. Experimental advances allow phenomena on this scale to be probed in unprecedented detail \cite{rondelezATP2005,alexanderBustamante2019,wruck2017}, but determining precisely how specific processes work and the role of evolutionary optimization remains a major challenge.
And while impressive progress has already been made engineering synthetic DNA \cite{Rothemund2006,SeemanSleiman2017} and protein \cite{HuangBaker2016,korendovychNovoProteinDesign2020} structures, we do not understand how to design \emph{de novo} nanomachines for nonequilibrium environments well enough for nanotechnology to rival the complexity of cellular machines.

For a microscopic system evolving out of equilibrium, thermodynamic quantities like entropy, heat, and work can be meaningfully defined only at the level of individual trajectories \cite{Seifert2012}. An ensemble of trajectories has distributions of thermodynamic properties with forms that depend on the system's non-equilibrium evolution \cite{Seifert2012,SpinneyFord2012}. This immediately suggests an optimization problem whereby a protocol $\bm{\lambda} (t)$ drives a system between given initial and final states to produce a desired distribution of thermodynamic properties. A common aim is to minimize average dissipated work. This is important for optimal bit erasure \cite{ZulkowskiDeweese2015,proesmansPRL2020,proesmansPRE2020}, as well as experimental measurements of the equilibrium free energy of biomolecules from nonequilibrium force pulling experiments and simulations~\cite{DellagoHummer2013,SpinneyFord2012,YungerHalpernJarzynski2016}. 
%Inverting the resulting force-displacement trajectories relies on rare “violations” of the second law of thermodynamics through the Jarzynski inequality [ref Jarzynski 2006]. Since errors in these methods scale with the exponential of the average work dissipated in a nonequilibrium process, protocols that minimize dissipative work are particularly important. 
Other targets include protocols that maximize thermodynamic efficiency for synthetic~\cite{Pietzonka2019,saha_infoengine_2021} and biological~\cite{lathouwersNonequilibriumEnergyTransduction2020} nanoengines and protocols minimize dissipated heat, for example in bit flipping operations~\cite{rotskoffOptimalControlNonequilibrium2015,rotskoffGeometricApproachOptimal2017}.

Modelling nonequilibrium processes is notoriously difficult, even when the equations of motion are precisely known.
A general method to optimize nonequilibrium driving protocols valid for systems of any complexity evolving arbitrarily far from equilibrium has heretofore not been elucidated. 
Existing work calculating optimal protocols has been limited to either extremely simple systems such a Brownian particle diffusing in a harmonic well or a single quantum dot~\cite{gomez-marinOptimalProtocolsMinimal2008,aurellOptimalProtocolsOptimal2011,schmiedlOptimalFiniteTimeProcesses2007,solonPhaseTransitionProtocols2018}, or applies only in the near-equilibrium regime~\cite{zulkowskiGeometryThermodynamicControl2012,bonancaOptimalDrivingIsothermal2014,largeOptimalDiscreteControl2019}. 
%An example of the latter  is theoretical work on optimal protocols for barrier crossing problems [Crooks, Sivak etc], recently been validated experimentally in DNA hairpin experiments [Tafoya 2019]. 
The assumption of near-equilibrium evolution restricts optimal protocols to free energy landscapes with low energy barriers, ruling out  most systems of interest, such as RNA molecules with pseudoknots, proteins, and biomolecular motors like ATP synthase. 

Inspired by recent computational advances in the machine learning community, we propose a method for computing optimal nonequilibrium protocols that is valid for complex systems evolving far-from-equilibrium. In particular, we leverage automatic differentiation (AD)~\cite{BrysonDenhamAD1962,brysonbook1975,rumelhartBackprop1986,BaydinAD2018}, a technique for computing gradients that repeatedly applies the chain rule to elementary computational steps. AD optimization has been recently applied in a range of scientific contexts, from quantum devices to self-assembly~\cite{vargas-hernandezFullyDifferentiableOptimization2021,CarlElla2021}. Using efficient AD algorithms developed in the context of training neural networks~\cite{tensorflow2015-whitepaper,NEURIPS2019_bdbca288,jax2018github} in conjunction with sophisticated GPU (graphical processing units) and TPU (tensor processing units) hardware, we compute gradients by \emph{backpropagating} through entire simulations, allowing us to find optimal protocols via gradient descent for a variety of systems. To illustrate the potential of this method, we here consider three canonical examples from the optimal protocol literature and drive evolution much farther out of equilibrium than previously possible. First, we consider Monte Carlo (MC) simulations and use AD to derive optimal protocols for flipping the magnetization of a 2D Ising lattice. The AD protocols perform similarly to existing near-equilibrium theoretical results in the linear regime and outperform the near-equilibrium theory in the far-from-equilibrium regime. Next, we treat molecular dynamics (MD) simulations,  reproducing classic analytical results for a single Brownian particle in a time-varying harmonic potential. Finally, we examine barrier-crossing on a double-well potential driven by a moving harmonic potential, a problem that maps onto biomolecular unfolding processes~\cite{sivakThermodynamicGeometryMinimumdissipation2016,tafoyaUsingSystemEquilibrium2019}. After recreating existing results, we probe the far-from-equilibrium regime of barrier crossing, demonstrating the capability of our method to capture departures from the near-equilibrium optimal protocols.

\section{Results}
\subsection{Differentiation of MC simulations: Nanomagnetic spin systems}

Computers dissipate large amounts of heat when performing logical operations via bit erasure, which reverses nanomagnetic spins~\cite{lambsonExploringThermodynamicLimits2011,hongExperimentalTestLandauer2016}. This has motivated recent studies investigating minimum-dissipation protocols for magnetization reversal with the 2D Ising model~\cite{rotskoffOptimalControlNonequilibrium2015,gingrichNearoptimalProtocolsComplex2016,rotskoffGeometricApproachOptimal2017}. 
%, a canonical workhorse in statistical mechanics that has uses far beyond the study of ferromagnetism, from DNA condensation [Vtyurina2016] to neural networks [Hopfield]. 
%The ubiquity
%vast utility and nonlinear dynamics
%of the Ising model makes it a good case study for non-equilibrium protocol optimization.
The system is described by the Hamiltonian
\begin{equation}
    H(B(t)) = -\sum_{\langle i,j \rangle} J_{ij}\sigma_i\sigma_j - B(t)\sum_{i} \sigma_i,
    \label{eq:ising_hamiltonian}
\end{equation}
where $\sigma_i = \pm1$ are the spins, $\langle ij \rangle$ indicates a sum over all nearest neighbour spins, $J_{ij}$ is the coupling between spins $i$ and $j$, and $B(t)$ is the (time-dependent) external magnetic field.

In the linear response (near-equilibrium) regime, Crooks and co-workers developed a general formalism for computing optimal protocols based on thermodynamic geometry~\cite{sivakThermodynamicMetricsOptimal2012a,zulkowskiGeometryThermodynamicControl2012}. Rotskoff and Crooks applied this theory to yield the theoretical optimal protocol for varying external magnetic field $B$ and spin-spin coupling strength $J$ (equivalent to varying temperature) to reverse magnetization on an Ising lattice~\cite{rotskoffOptimalControlNonequilibrium2015}. More recently, Gingrich et al.~\cite{gingrichNearoptimalProtocolsComplex2016} explored the same problem using a numeric approach, in which the space of low dissipation protocols is explored with a Monte Carlo scheme. This yields a number of degenerate, near-optimal protocols, but like the work of Rotskoff and Crooks~\cite{rotskoffOptimalControlNonequilibrium2015} is limited by the assumption of near-equilibrium evolution.

Inspired by previous work, we examine the non-equilibrium magnetization reversal of a 2D periodic lattice of spins driven by a protocol that varies both magnetic field $B(t)$ and temperature $T(t)$: $\bm{\lambda}(t) = (B(t), T(t))$, but push evolution beyond the near-equilibrium regime, benchmarking against the linear response formalism of Rotskoff and Crooks~\cite{rotskoffOptimalControlNonequilibrium2015}.

We seek to minimize  the total entropy production $\Delta S$,
%(which \citet{Gingrich2016} call `dissipation' $\omega$)
a proxy for the heat dissipated to the environment during the ``bit flip'' that quantifies the irreversibility of the process~\cite{SpinneyFord2012,jarzynskiNonequilibriumWorkRelations2008,crooksExcursionsStatisticalDynamics1999,crooksNonequilibriumMeasurementsFree1998,gingrichNearoptimalProtocolsComplex2016}:
\begin{equation}
    \Delta S = k_B\ln{\frac{P^F[\bm{x}]}{P^R[\bm{\bar{x}}]}}.
    \label{eq:irreversibility}
\end{equation}
Here $P^F[\bm{x}]$ is the probability of observing a particular trajectory $\bm{x}$ during the forward evolution of a system, and $P^R[\bm{\bar{x}}]$ is the probability of observing the exact time-reversal of that trajectory, $\bm{\bar{x}}$. 
%Using Crooks' convention [ref], 

To find the optimal protocol for varying $B(t)$ and $T(t)$, we have written a Monte Carlo simulator using JAX~\cite{jax2018github}, a python library with built-in automatic differentiation and just-in-time compilation. The code carries out standard Glauber dynamics~\cite{glauberTimeDependentStatistics1963} and iteratively updates the grid points with even then odd lattice index $i$ using the spin flip probability

\begin{equation}
    P(\sigma_i \rightarrow -\sigma_i) = \frac{1}{1 + e^{\beta\Delta E_i}},
    \label{eq:glauberP}
\end{equation}
where $\beta=1/k_BT$ is the usual inverse thermal energy and the change in lattice energy resulting from the flip of spin $\sigma_i$ can be computed using the sum of its nearest neighbour spins $\sum\limits_{j} \sigma_j$:
\begin{equation}
    \Delta E_i = 2\sigma_i(J_{ij}\sum\limits_{j} \sigma_j - B)
    \label{eq:deltaE}
\end{equation}

Our code compiles to run rapidly on GPUs or TPUs and is differentiable: given a Monte Carlo trajectory of spins under some protocol $\bm{\lambda}(t)$, we can compute the gradient of any function of the trajectory with respect the protocol. This gradient can be computed in either forward mode or reverse mode. In forward mode, the computational work for computing the gradient depends on the number of parameters characterizing the protocol $\bm{\lambda}(t)$. In reverse mode, the computational work is {\sl independent} of the number of parameters but it is necessary to hold the entire trajectory and all intermediate derivatives in memory, which can be a significant constraint.

Here, the parameterizations we use for $\bm{\lambda}(t)$ are low dimensional relative to the length of the trajectories considered, so gradients are computed with forward mode differentiation. We have found it most convenient to
parameterize the protocol using a Chebyshev polynomial basis
\begin{equation}
    \lambda(t) = \theta_0 T_0(t) + \theta_1 T_1(t) + \ldots + \theta_k T_k(t),
\end{equation}
where the $T_i$ is the $i^{th}$ Chebyshev polynomial and $k$ is a hyperparameter (typically we choose $k=32$), but many other parameterizations are possible.  

The loss function (\ref{eq:irreversibility}) is computed for our trajectories as follows. Each time step in the forward evolution can be formulated as a sequence of two sub-steps: first, the external protocol parameters are updated (this is where external work is performed), and then the system performs a transition to a new microstate (this is where heat is exchanged with the bath)~\cite{jarzynskiNonequilibriumWorkRelations2008,crooksExcursionsStatisticalDynamics1999,crooksNonequilibriumMeasurementsFree1998}.
%\begin{equation}
%    (x_n, \lambda_n) \rightarrow (x_n, \lambda_{n+1}) %\rightarrow (x_{n+1}, \lambda_{n+1}) \ldots
%\end{equation}
%The time-reversed process reverses these sub-steps: 
%the system first takes a random step to a new microstate, then the external protocol parameters are updated:
%\begin{equation}
%    (x_{n+1}, \lambda_{n+1}) \rightarrow (x_n, %\lambda_{n+1}) \rightarrow (x_n, \lambda_n) \ldots
%\end{equation}
If $\rho_A(x_0)$ is the probability of drawing microstate $x_0$ from the equilibrium distribution corresponding to state $A$, then
the probability of observing the forward trajectory is given by 
\begin{equation}
    P^F[\bm{x}] = \rho_A(x_0) P(x_0 \rightarrow x_1; \bm{\lambda}_1) \ldots P(x_{N-1} \rightarrow x_N; \bm{\lambda}_{N})
    \label{eq:pf}
\end{equation}
where  $P(x_{i-1} \rightarrow x_i; \bm{\lambda}_i)$ is the transition probability between lattice states $x_{i-1}$ and $x_i$ at protocol parameter values $\bm{\lambda}_i$. Correspondingly, the  probability of observing the time-reversed trajectory $\bm{\bar{x}}$ is
\begin{equation}
    P^R[\bm{\bar{x}}] = \rho_B(x_N) P(x_N \rightarrow x_{N-1}; \bm{\lambda}_N) \ldots P(x_1 \rightarrow x_0; \bm{\lambda}_1)
    \label{eq:pr}
\end{equation}

Formulating the evolution as a succession of accepted and rejected spin flips and noting that for the Glauber transition probability (\ref{eq:glauberP}), $1 - P(\Delta E) = P(-\Delta E)$, with $\Delta E$ is given by (\ref{eq:deltaE}), we can combine (\ref{eq:pf}) and (\ref{eq:pr}) to obtain:

\begin{equation}
    \frac{P^F[\bm{x}]}{P^R[\bm{\bar{x}}]} = \frac{\rho_A(x_0) \prod_{i}P(\Delta E_i)\prod_{j}P(-\Delta E_j)}{\rho_B(x_N) \prod_{i}P(-\Delta E_i)\prod_{j}P(-\Delta E_j)}.
    \label{eq:pratio}
\end{equation}

Here, the products containing $i$ include all spins that flipped successfully and those containing $j$ are failed spin flips. Note that the probability for a spin \textit{not} to flip is the same in the forward and reverse trajectories, while the $\Delta E$ terms for accepted spin flips along the forward and reverse trajectory differ in sign. The ratio of probabilities of drawing the initial and final states from their respective equilibrium ensembles, $\frac{\rho_A(x_N)}{\rho_B(x_0)}$, is given by 
\begin{equation}
    \frac{\rho_A(x_N)}{\rho_B(x_0)} = \frac{Z_B}{Z_A}\frac{e^{-\beta_0H(x_0, \lambda_0)}}{e^{-\beta_NH(x_N, \lambda_N)}},
\end{equation}
where $Z_B$/$Z_A=1$ since the magnitude of the external field is identical (and equal to 1) in the initial and final states of our simulations.

Plugging (\ref{eq:glauberP}) into (\ref{eq:pratio}) and rearranging (\ref{eq:irreversibility}), we find 
%after some algebra 
that the dissipation in our simulations is given by
\begin{equation}
    \frac{\Delta S}{k_B} \equiv \omega = \beta_N H(x_N, \lambda_N) -  \beta_0 H(x_0, \lambda_0)  - \sum_i \beta_i \Delta E_i,
    \label{eq:loss}
\end{equation}
where 
%$H(x_0, \lambda_0)$ and $H(x_N, \lambda_N)$ are the Hamiltonians for the system in the initial and final states, and 
we take a sum of the system energy changes $\Delta E_i$ following successful spin flips $i$, multiplied by the inverse temperature $\beta_i$ at which the flip occurred.

We use JAX's automatic differentiation to compute the gradient $\nabla_{\bm{\theta}} \langle \omega \rangle$ of this dissipation. Due to the discrete choices inherent in Monte Carlo simulations, care must exercised in computing gradients. In particular, the dissipation depends on external parameters through discrete spin flip operations -- dictated by a Metropolis-like acceptance criterion -- which are not themselves differentiable; a similar issue arises in the context of training stochastic neural networks~\cite{bengioEstimatingPropagatingGradients2013}. Instead of directly backpropagating through the loss function, we proceed as follows. The % dissipation as a functional of the system trajectory $\bm{x}(t)$ and the external parameters $\bm{\theta}$ and its 
average dissipation over all possible trajectories $\bm{x}(t)$ for external protocol parameters $\bm{\theta}$ is given by (see also Ref. \cite{geigerOptimumProtocolFastswitching2010}):
\begin{equation}
    \langle \omega(\bm{\theta})\rangle = \int \mathcal{D}\bm{x}(t) P[\bm{x}(t); \bm{\theta}] \omega[\bm{x}(t), \bm{\theta}],
    \label{eq:avg_omega}
\end{equation}
where $\int \mathcal{D}\bm{x}(t)$ is an integration over all possible trajectories, $P[\bm{x}(t); \bm{\theta}]$ is the probability weight associated with each trajectory, and $\omega[\bm{x}(t), \bm{\theta}]$ is the dissipation (total entropy production) for each trajectory. 

Applying the product rule and noting that $\nabla P = P \nabla\ln{P}$, the gradient of $\langle \omega(\bm{\theta})\rangle$ is 
\begin{equation}
    \nabla_{\bm{\theta}} \langle \omega(\bm{\theta})\rangle = \langle (\nabla_{\bm{\theta}} \ln P) \omega \rangle + \langle \nabla_{\bm{\theta}} \omega \rangle.
\label{eq:reinforce}
\end{equation}
Note that merely averaging the gradient $\nabla_{\bm{\theta}} \omega$ over a batch of simulated trajectories does {\sl not} yield the correct average, since the probability of observing a trajectory is itself dependent on protocol parameters $\bm{\theta}$. A similar approach to finding stochastic gradients is used for taking gradients in the REINFORCE algorithm~\cite{williamsSimpleStatisticalGradientfollowing1992}, widely used in reinforcement learning.

We carry out $N$ Monte Carlo simulations of the 2D Ising lattice evolving under assumed protocol $\bm{\lambda}(\bm{\theta})$. For each trajectory, we compute the gradients $\nabla_{\bm{\theta}}\ln{P}$ and $\nabla_{\bm{\theta}}\omega$. Plugging these into Equation (\ref{eq:reinforce}) and averaging gives us a Monte Carlo estimate of the gradient of interest, $\nabla_{\bm{\theta}}\langle\omega(\theta)\rangle$.
We then use the Adam optimizer~\cite{kingmaAdamMethodStochastic2017} to minimize the loss (Eq. \ref{eq:loss}).

%in the Rotskoff 2015: 128x128 lattice, two trials with temp constrained to 3, 4, then one unconstrained
\begin{figure}
    \centering
    \includegraphics[scale=0.4]{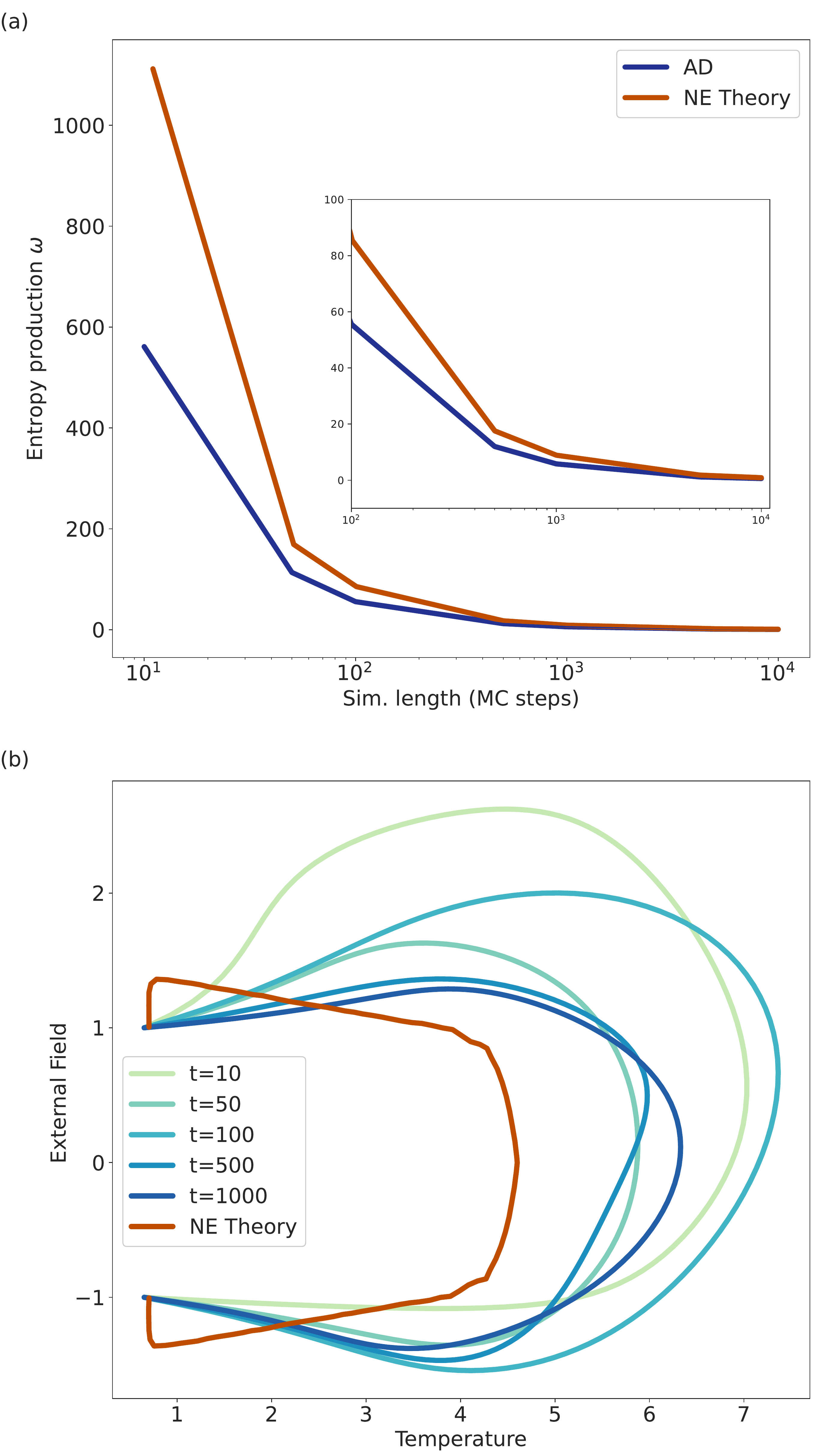}
    \caption{(A) Minimum entropy dissipation for different simulation lengths on a 32x32 lattice, averaged over n=2560 trajectories (standard errors of the mean are smaller than the line widths). The AD protocols in each case were derived using the Adam optimizer~\cite{kingmaAdamMethodStochastic2017} with gradients averaged over batches of N=256. AD protocols are able to outperform the optimal entropy production that results from using the near-equilibrium (NE) theoretical protocol of Rotskoff and Crooks~\cite{rotskoffOptimalControlNonequilibrium2015} and approach the minimum dissipation value of Rotskoff and Crooks~\cite{rotskoffOptimalControlNonequilibrium2015} as the near-equilibrium regime is approached. Inset: zoom-in of near-equilibrium region. (B) Optimal protocols for reversing magnetization for five simulation lengths: t=10, 50, 100, 500, and 1000. Forward evolution in time occurs along the counterclockwise direction. Curves for t=5000 and t=10000 are nearly identical to the t=1000 curve and are not shown. That the AD curves do not converge to the shape of the near-equilibrium (NE) theoretical curve, yet produce lower dissipation in (A), suggests a flat loss landscape in the optimal region in the near-equilibrium regime, in accordance with previous findings~\cite{rotskoffOptimalControlNonequilibrium2015,gingrichNearoptimalProtocolsComplex2016}.}
    \label{fig:ising}
\end{figure}

Figure~\ref{fig:ising} (A) shows our lowest achieved average entropy production $\langle \omega \rangle$ from AD protocols alongside $\langle \omega \rangle$ for the near-equilibrium theoretical protocol of Rotskoff and Crooks~\cite{rotskoffOptimalControlNonequilibrium2015} for seven different simulation lengths on a 32x32 lattice ($\tau$=10, 50, 100, 500, 1000, 5000, and 10000 MC time steps). The longer the simulation, the closer the magnetization reversal is to quasi-static. $\langle \omega \rangle$ values are averages over n=2560 trajectories. In all cases, AD outperforms the near-equilibrium theory.%, and appears to be converging to the near-equilibrium theoretical protocol's $\langle \omega \rangle$ curve as the near-equilibrium regime is approached. 
We repeated the t=50, t=100, t=500, and t=1000 simulations on multiple lattice sizes up to 512x512 and found the normalized entropy production on the 32x32 lattice is within 1\% of its converged, infinite-lattice value; see SI Figure S3. The AD-derived protocols also outperformed the near-equilibrium theory regardless of lattice size. This suggests AD optimization can be effectively performed on smaller lattices to find optimal protocols for larger lattices.

Figure~\ref{fig:ising} (B) shows the optimal protocols corresponding to the simulation lengths in (A) along with the near-equilibrium theoretical curve of Rotskoff and Crooks~\cite{rotskoffOptimalControlNonequilibrium2015}. %For this lattice size, batches of N=256 are sufficient to yield a well-converged average entropy production. 
Curves for t=5000 and t=10000 are omitted as they were near-identical to the t=1000 curve. While the near-equilibrium theoretical protocol is necessarily time-symmetric~\cite{rotskoffOptimalControlNonequilibrium2015}, our protocols appear to break this symmetry; see SI Fig.~S5. Like the near-equilibrium theoretical result, the AD protocols avoid the critical phase transition region~\cite{rotskoffOptimalControlNonequilibrium2015}, but they do not appear to be converging to the exact shape of the near-equilibrium theoretical curve as equilibrium is approached. We also observed a flat loss function for t=5000 and t=10000 over 1000 optimization iterations, as shown in SI Figure S2. The fact that our curves perform comparably to the near-equilibrium theoretical curve in the near-equilibrium limit, but are differently shaped,  %As we pushed our simulation length farther into the near-equilibrium regime ($\tau \sim$5000 MC steps), we find multiple AD curves that perform similarly to the theoretical optimal curve of Rotskoff and Crooks~\cite{rotskoffOptimalControlNonequilibrium2015} \textcolor{red}{(CONFIRM THIS AGAIN)}. In particular, even when using the theoretical result from Rotskoff and Crooks~\cite{rotskoffOptimalControlNonequilibrium2015} as an initial guess, different optimal protocols are found that yield comparable entropy production \textcolor{red}{(CONFIRM AGAIN)}. 
suggests that the entropy dissipation landscape is relatively flat in the region of optimal protocols, and that there is a degenerate space of nearly-optimal solutions. Indeed, all of the AD protocols in Fig.~\ref{fig:ising} (B) perform comparably well at longer simulation lengths; see SI Figure S6. This is in agreement with the findings of Gingrich et. al.~\cite{gingrichNearoptimalProtocolsComplex2016} and noted by Rotskoff and Crooks~\cite{rotskoffOptimalControlNonequilibrium2015}, who predict `weakly constrained' protocols in the non-critical region of $(B(t), T(t))$-space.

%from Sivak Crooks 2016: "Under the linear response approximation, the shape of the optimal protocol is not a function of the allocated time interval."

\subsection{Optimal Protocols for Brownian Dynamics}
We now consider the  molecular dynamics (MD) of isothermal evolution of Brownian particles , where the total entropy production is equal to the dissipated external work ($W_D$)~\cite{SpinneyFord2012}, which we use as our loss function in the following case studies.% Optimal protocols in the context of the subsequent sections should be understood to mean those that minimize $W_D$ under given constraints.

\subsubsection{Brownian particle in a harmonic potential}

Some of the earliest work identifying optimal non-equilibrium protocols focused on the paradigmatic colloidal Brownian particle in a harmonic trap~\cite{dekoningOptimizingDrivingFunction2005,schmiedlOptimalFiniteTimeProcesses2007,gomez-marinOptimalProtocolsMinimal2008,geigerOptimumProtocolFastswitching2010,Seifert2012}. %When forced by a harmonic potential, this represents optical tweezer experiments~\cite{sivakThermodynamicGeometryMinimumdissipation2016}.  
Exact optimal protocols for varying the stiffness of the center and stiffness of the harmonic potential were found by Schmiedl and Seifert~\cite{schmiedlOptimalFiniteTimeProcesses2007} using
variational calculus. Strikingly, the solutions have discrete `jumps' in the parameters ~\cite{thenComputingOptimalProtocol2008,geigerOptimumProtocolFastswitching2010}. We note that approaches based on the linear response approximation are incapable of discovering these jumps since they assume protocols are differentiable~\cite{dekoningOptimizingDrivingFunction2005,sivakThermodynamicMetricsOptimal2012a}.

We reproduce these early results by using JAX-MD~\cite{jaxmd2020} to automatically differentiate molecular dynamics simulations of a colloid subjected to the moving harmonic potential 
\begin{equation}
    V(x,t) = \frac{1}{2}(x - \lambda(t))^2,
    \label{eq:potential_movingtrap}
\end{equation} 
where here $\lambda(t)$ is the time dependent position of the trap. We seek a protocol that 
%We use a customized version of the Brownian dynamics integrator included with JAX, MD that enables us to keep track of the log probability of a particular thermal `kick' in order to apply equation (\ref{eq:reinforce}) in calculating gradients. This is necessary because, similarly to MC dynamics, Langevin evolution is stochastic (due to the random forcing term) and the derivatives of expectation values of the loss functions must be determined via the REINFORCE-type method described above.
that minimizes the total work dissipated in moving the trap from $\lambda_i$ to $\lambda_f$. As before, and following Crooks~\cite{crooksNonequilibriumMeasurementsFree1998}, evolution can be formulated as proceeding in two stages: (i) the external protocol is updated and then (ii) the particle makes a random transition to a new state. External work is done in the first step, implying:
\begin{equation}
    W =\sum_{n=0}^{N-1} [V(x_n; \lambda_{n+1})-V(x_n; \lambda_{n})].
    \label{eq:work}
\end{equation}
Since the free energy is the same in the initial and final ensembles, this external work is equivalent to the `dissipated work' $W_D$.

We performed Brownian dynamics simulations using one of the sets of parameters considered by Schmiedl and Seifert~\cite{schmiedlOptimalFiniteTimeProcesses2007}: $\Delta \lambda = \lambda_f - \lambda_i = 5$; $k_BT = \mu = 1$, with $k_BT$ the usual thermal energy and $\mu$ the mobility of the colloid, and total simulation time of $t_f = 2.69$ units, the time that theoretically yields the highest ratio between the work dissipated by a `naive', linear trap protocol and the dissipated work corresponding to the optimal protocol~\cite{schmiedlOptimalFiniteTimeProcesses2007}. 

To parameterize our protocols, we consider piecewise linear $\lambda(t)$, specified by the values at 8 distinct time points.
%; this enables us to capture discrete jumps in the protocol. 
Starting from an initial guess of a linear protocol, we perform optimization with Adam~\cite{kingmaAdamMethodStochastic2017} on batches of N = 5000 MD simulations with learning rate $\alpha = 0.1$, integration time step $dt = 0.001$, and an initial equilibration period of $0.1$ simulation time units prior to trap motion, we are able to reproduce the exact theoretical optimal curve derived by Schmiedl and Seifert~\cite{schmiedlOptimalFiniteTimeProcesses2007} %\begin{equation}
    %\lambda(t) = \lambda_f\frac{(t+1)}{(t_f + 2)} + \lambda_i
%\end{equation}
within 100 optimization iterations, taking a few minutes on a GPU. Our calculation reproduces the theoretical ratio between the work dissipated by the optimal and linear protocols, $W_{lin}/W_{opt} \approx 1.14$ for $t_f = 2.69$~\cite{schmiedlOptimalFiniteTimeProcesses2007}.

\begin{figure}
    \centering
    \includegraphics[width=\columnwidth]{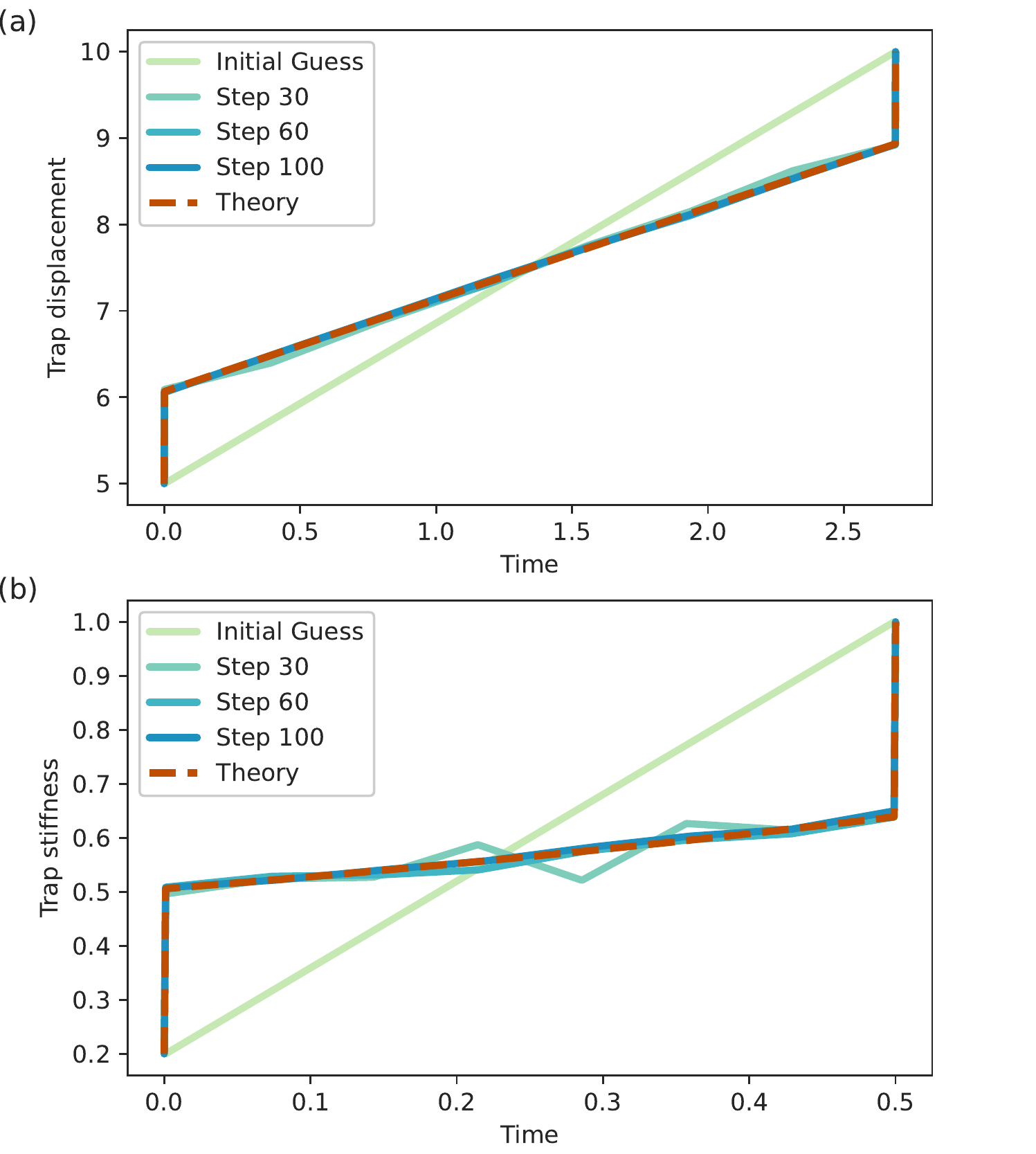}
    \caption{Automatic differentiation-based optimization rapidly converges to the exact theoretical optimal protocol for the cases of a Brownian particle in (A) a moving harmonic trap and (B) a trap with time-varying stiffness. The discrete jumps at the beginning and the end of the protocol found originally by Schmiedl and Seifert~\cite{schmiedlOptimalFiniteTimeProcesses2007} are recaptured by our method. 
    }
    \label{fig:browniantrap}
\end{figure}

Our methodology also successfully reproduces the exact theoretical optimal protocol for varying the stiffness of a harmonic potential 
$V(x,t) = \lambda(t)x^2/2$ within 100 optimization iterations; see Materials \& Methods for details.
%which is, according to \citet{Seifert2007},
%\begin{equation}
%    \lambda(t) = \frac{\lambda_i - c(1 + ct)}{(1+ct)^2}
%\end{equation}
%where
%\begin{equation}
%    c = \frac{-1-\lambda_f t_f + \sqrt{1 + 2\lambda_it_f + %\lambda_f\lambda_it_f^2}}{t_f(2 + \lambda_ft_f)}
%\end{equation}
Figure~\ref{fig:browniantrap} summarizes our results.

\subsubsection{Driven barrier crossing}
\begin{figure*}
    \includegraphics[scale=0.4]{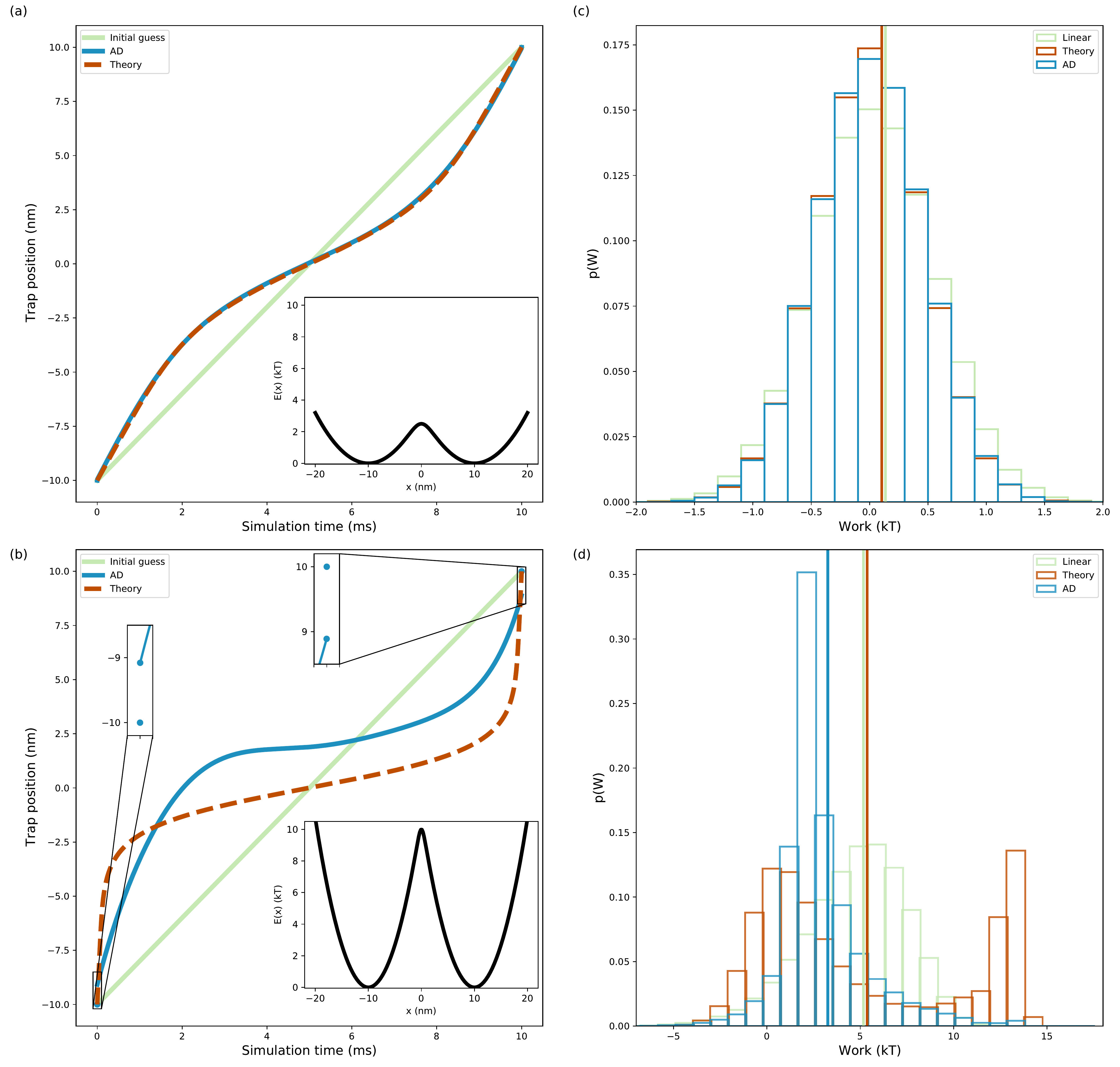}
    \caption{(A) The automatic differentiation-based optimal protocol converges to the near-equilibrium (NE) theoretical result of Sivak and Crooks~\cite{sivakThermodynamicGeometryMinimumdissipation2016} in the near-equilibrium regime, corresponding to a 2.5~\si{\kT} barrier landscape. Inset: a 2.5~\si{\kT} barrier energy landscape. (B) The AD protocol is asymmetric and differs from the near-equilibrium (NE) theoretical result of Sivak and Crooks~\cite{sivakThermodynamicGeometryMinimumdissipation2016} in the farther-from-equilibrium regime represented by a 10~\si{\kT} barrier landscape. Upper right insets reveal a discrete jump at the beginning and end of the optimal protocol, in agreement with results for similar systems~\cite{schmiedlOptimalFiniteTimeProcesses2007,gomez-marinOptimalProtocolsMinimal2008,thenComputingOptimalProtocol2008,espositoFinitetimeThermodynamicsSinglelevel2010}. Lower right inset: a 10~\si{\kT} barrier energy landscape. (C) Probability work distributions associated with the protocols in (A). AD ($\langle W_D \rangle = 0.104 \pm 0.001~\si{\kT}$) performs as well as the near-equilibrium (NE) theory ($\langle W_D \rangle = 0.104 \pm 0.001~\si{\kT}$) within error, and both outperform a naive linear protocol ($\langle W_D \rangle = 0.136 \pm 0.002~\si{\kT}$). (D) Probability work distributions for the protocols shown in (C). Here the AD protocol ($\langle W_D \rangle = 3.260 \pm 0.007~\si{\kT}$) significantly outperforms the near-equilibrium (NE) theory ($\langle W_D \rangle = 5.37 \pm 0.02~\si{\kT}$), which fails to unfold all particles and is thus also beaten by a naive linear protocol ($\langle W_D \rangle = 5.172 \pm 0.009~\si{\kT}$).}
    \label{fig:barrier_crossing}
\end{figure*}

We now turn to the more complex situation of driven Brownian motion on a bistable potential, a model used widely in soft matter to represent biomolecular unfolding via AFM or optical tweezers~\cite{dudkoIntrinsicRatesActivation2006c,woodsideReconstructingFoldingEnergy2014c}, as well as optimal protocols for bit erasure~\cite{aurellRefinedSecondLaw2012,proesmansPRE2020}. %Because of its comparative tractability combined with relatability to experiments, the model is an appealing target for optimal protocol investigations, inspiring 
Sivak and Crooks~\cite{sivakThermodynamicGeometryMinimumdissipation2016} consider a Brownian particle driven across a bistable potential energy landscape (see Figure \ref{fig:barrier_crossing} insets and Materials \& Methods for details of the potential) by a harmonic trap with a time dependent minimum $\lambda(t)$. The trap drives barrier crossing from one minimum to the other.

Following References \cite{sivakThermodynamicGeometryMinimumdissipation2016} and \cite{tafoyaUsingSystemEquilibrium2019}, we performed molecular dynamics simulations of barrier crossing using parameters approximating DNA hairpin unfolding experiments with optical tweezers; see Materials \& Methods for details. For quantitative comparison with previous work, we make the same simplifying assumptions as Sivak and Crooks \cite{sivakThermodynamicGeometryMinimumdissipation2016}: (i) the free energies of the initial and final equilibrium states are equal and (ii) the two landscape wells have equal curvature. Note that our method does not require these assumptions. %\citet{Tafoya2019} study 2 hairpins experimentally, one of whose dynamics falls in the near-equilibrium regime and the other whose dynamics are too slow to be well described by the \citet{SivakCrooks2016} theory~\cite{Tafoya2019}. 
We proceed by calculating dissipated work with Eq. \ref{eq:work}. We consider two free energy landscapes, with barrier heights 2.5~\si{\kT} and 10~\si{\kT}, corresponding to the near-equilibrium regime and a farther-from-equilibrium regime, respectively; SI Fig.~S7 contains details of how we quantified distance from equilibrium. Here, a barrier height of 2.5~\si{\kT} (10~\si{\kT}) maps roughly onto the unfolding of a 6 (20) base pair DNA hairpin~\cite{Pfitzner2013,engel2019dna,woodsideDirectMeasurementFull2006}. 

%Avg work for 2kT linear is:  0.13632213 +/- 0.0016529334
%Avg work for 2kT AD is:  0.104464024 +/- 0.0014616443
%Avg work for 2kT theory is:  0.10363535 +/- 0.0014637125
%Avg work for 10kT linear is:  5.1721287 +/- 0.008619704
%Avg work for 10kT AD is:  3.2601855 +/- 0.006573348
%Avg work for 10kT theory is:  5.366506 +/- 0.016962543

Figure~\ref{fig:barrier_crossing} presents the results of using automatic differentiation-based optimization to find optimal trap protocols $\lambda(t)$ for driven barrier crossing; landscape profiles are shown as insets. In the near-equilibrium regime (barrier height 2.5~\si{\kT}), optimizing over a batch of N = 2504 trajectories, our method converges to the near-equilibrium theoretical result of Sivak and Crooks~\cite{sivakThermodynamicGeometryMinimumdissipation2016} after 1000 optimization iterations, with most of the convergence achieved after a couple hundred optimization steps (taking a few hours on TPU). The shape of the optimal protocol -- faster trap motion at the beginning and ends of the motion and a slowing down in the central barrier region -- reflects the fact that the minimal work is dissipated if the trap slows down in the vicinity of the barrier to `wait'
for the system to harness thermal energy kicks to surmount it~\cite{sivakThermodynamicGeometryMinimumdissipation2016,tafoyaUsingSystemEquilibrium2019}.

We compared the limiting probability distributions (across $\sim$1e5 MD simulations) of dissipated work for the near-equilibrium theoretical protocol, our result, and a naive linear protocol.
%and after 10,000 trajectories, which we take to represent the limiting work distribution. 
The AD-optimized and near-equilibrium theoretical distributions agree within error, each having a mean work of $\langle W_D \rangle = 0.104 \pm 0.001~\si{\kT}$. Both protocols outperform the naive linear protocol, which gives  $\langle W_D \rangle = 0.136 \pm 0.002~\si{\kT}$. Errors are standard errors of the mean.

The AD-based optimization allows us to probe far beyond the
 near-equilibrium regime (Figure~\ref{fig:barrier_crossing} (C) and (D)). With a 10~\si{\kT} barrier landscape, the automatic differentiation-based optimal protocol outperforms the near-equilibrium theory. Here, our algorithm finds that a non-symmetric protocol is optimal, whereas linear theory necessarily predicts that it is symmetric~\cite{sivakThermodynamicMetricsOptimal2012a}. Intuitively, the trap needs to spend more time in the vicinity of the barrier to successfully `drag' the particle along: the bimodal distribution of the near-equilibrium theoretical $p(W)$ in Figure~\ref{fig:barrier_crossing} (D) reveals that not every particle successfully `unfolds' under the near-equilibrium optimal protocol: some are left behind in the `folded' state after the trap has completed its motion. These trajectories maximize the external dissipated work: after the trap stops moving, work can no longer accrue according to \ref{eq:work}, even if the particle eventually hops to the unfolded well. In contrast,  AD-based optimization finds a protocol that `unfolds' all molecules in simulation time, leading to a significantly lower average dissipated work of $\langle W_D \rangle = 3.260 \pm 0.007~\si{\kT}$ compared to the near-equilibrium theoretical mean work of $\langle W_D \rangle = 5.37 \pm 0.02~\si{\kT}$. Here, a naive linear protocol ($\langle W_D \rangle = 5.172 \pm 0.009~\si{\kT}$) also outperforms the near-equilibrium theory.
 
 The AD-based $\lambda(t)$ features discrete jumps at the beginning and end of the protocol that are absent in the linear response optimum (see upper insets of Figure~\ref{fig:barrier_crossing} (C)). Discrete jumps have been observed in multiple other studies of minimum-dissipation protocols~\cite{gomez-marinOptimalProtocolsMinimal2008,thenComputingOptimalProtocol2008,espositoFinitetimeThermodynamicsSinglelevel2010}, and indeed were posited by Schmiedl and Seifert~\cite{schmiedlOptimalFiniteTimeProcesses2007} to be a ``generic feature of the optimal protocol for arbitrary potentials.'' Recent work corroborates the universality of jump features in optimal protocols~\cite{blaberStepsMinimizeDissipation2021}. Since the near-equilibrium theory assumes protocols to be differentiable, it necessarily miss these features~\cite{sivakThermodynamicMetricsOptimal2012a,zulkowskiGeometryThermodynamicControl2012,ZulkowskiDeweese2015}. 
%Because loss landscapes may be non-convex in the far-from-equilibrium regime, the optimal protocols identified by AD are not necessarily unique.

We note that although here we have focused here on the form of landscape studied in previous literature, our method allows the user to perform a similar barrier crossing optimization for virtually any free energy landscape, such as bespoke free energy landscapes containing nontrivial features -- like intermediate states -- that map to complex biomolecules.

\section{Conclusion}

We have demonstrated the viability of automatic differentiation (AD) to identify optimal non-equilibrium protocols for both Monte Carlo and molecular dynamics simulations. The method performed as well as existing near-equilibrium theoretical results in the near-equilibrium regime for both magnetization reversal on a 2D Ising lattice and driven barrier crossing. Critically, the AD algorithm easily extends to far-from-equilibrium conditions, where it significantly outperforms existing near-equilibrium theoretical protocols.

Our work considers fixed simulation times with one- and two- dimensional protocols, though the framework is much more general than this, and essentially arbitrary constraints and protocol parameters are possible, e.g. multidimensional external protocols and arbitrary loss functions. For example, one could attempt to maximize the accuracy of Jarzynski-based free energy landscape reconstructions for a given amount of experimental or computational time; optimize the speed or efficiency of nanoengines; or minimize the time taken to unfold a molecule. Further, in the JAX-MD code suite, non-Brownian dynamics can easily be simulated, opening up the possibilities of optimizing protocols for systems like the recently-proposed active-matter based thermodynamic engine~\cite{Pietzonka2019}.

The main limitation of AD-based protocol optimization at present is the high computational cost of backpropagating gradients through entire simulations. We are currently exploring strategies to mitigate this, including importance sampling of trajectories. Improving performance will allow more experimentally-realistic models of complex systems to be studied. 

%in the far-from-eq'm regime, loss landscapes may not be convex, and the `optimal' protocols derived using AD are not necessarily unique.  

AD provides a valuable complement to existing near-equilibrium approaches to find optimal protocols, as it makes more complex systems and the far-from-equilibrium regime accessible. We are eager to see its manifold applications unfold in non-equilibrium protocol optimization and beyond.

\section{Materials and Methods}

\subsection*{2D Ising Model MC simulations}
We implement Glauber dynamics~\cite{glauberTimeDependentStatistics1963} on a 32x32 lattice with periodic boundary conditions using the JAX Python code suite~\cite{jax2018github}, and check our results using forward simulations on multiple other lattice sizes up to 512x512. 
Because the Hamiltonian (Eqn. \ref{eq:ising_hamiltonian}) contains only nearest-neighbour spin interactions, we use a checkerboard update scheme~\cite{preisGPUAcceleratedMonte2009} in which spin flips are proposed for all `even' spins and then all `odd' spins. Spin flips are accepted with the Glauber probability (Eqn. \ref{eq:glauberP}). 

As an initial protocol guess, we use a linear ramp for the external field and a quadratic curve for the temperature with fixed endpoints $(B(0), T(0))=(-1, 0.65)$ and $(B(\tau), T(\tau))=(1, 0.65)$; see SI for details. The exceptions are our t=5000 and t=10000 simulations, for which we use the t=1000 and t=5000 AD protocols as initial guesses, respectively. Each of the $B(t)$, $T(t)$ protocols are parametrized as 32-degree Chebyshev polynomials capturing the deviation of the final protocol from the initial guess.

Gradients are clipped at norm $=1$ and the Adam optimizer~\cite{kingmaAdamMethodStochastic2017} is used with $b1=0.9$, $b2=0.999$, $eps=1\times10^{-8}$, and a learning rate that begins at $\alpha=0.1$ and decays exponentially to $\alpha=0.01$ by the end of the optimization, which we run for 5000 iterations for the t=10, t=50, t=100, t=500, and t=1000 simulations. For the t=5000 and t=10000 simulations, the learning rate starts at $\alpha=0.01$ and decays exponentially to $\alpha=0.001$ over 1000 iterations. We compute gradients over batches of N=256 simulations. 

\subsection*{Molecular dynamics simulations}
We evolve a particle using the JAX-MD Python package~\cite{jaxmd2020} according to overdamped Langevin dynamics~\cite{carlonMolecularDynamicsSimulations}:

\begin{equation}
    \dot{r}(t) = \frac{F(r(t))}{m\gamma} - \sqrt{\frac{2k_BT}{m\gamma}}\eta(t)
    \label{eq:langevin}
\end{equation}

where $r(t)$ is the particle's position; $F(r(t))$ are the forces arising from the potential; $m$ is the particle mass; $\gamma$ is the friction coefficient; $k_BT$ is the usual thermal energy; and $\eta(t)$ is a random Gaussian-distributed noise term that simulates thermal coupling to the bath. 

\subsubsection*{Brownian particle in harmonic trap}
As noted in the text, we use $\mu = \frac{1}{m\gamma} = 1$, $kT = 1$, integration time step $\delta t = 0.001$, and run 100 steps with the Adam optimizer with with $b1=0.9$, $b2=0.999$, $eps=1\times10^{-8}$ and learning rate $\alpha=0.1$. Before beginning protocols, the colloid is equilibrated for $0.1$ simulation time units, until its energy stabilizes. For the moving trap simulations: we use batches of N=5000 trajectories to estimate gradients; run for a total simulation time of $t_f = 2.69$ simulation time units; set trap stiffness $k=1$; and use $\lambda_i = 5$ and $\lambda_f = 10$. For the varying stiffness simulations, we use batches of N=100 000 trajectories; run for a total simulation time of $t_f = 0.5$; and increase stiffness from $\lambda_i = 0.2$ to $\lambda_f = 1$. The average dissipated work is given by (\ref{eq:reinforce}); however, the first term is zero for our Brownian evolution, as the randomness originates in Gaussian noise generated by a uniform probability distribution. This is analogous to the `reparametrization trick' used in variational autoencoders~\cite{kingmaAutoEncodingVariationalBayes2014}. We therefore straightforwardly compute $\langle \nabla_{\bm{\theta}} \omega \rangle$ in our optimization.

\subsubsection*{Driven barrier crossing}
We simulate Brownian motion on the bistable landscape 
\begin{equation}
    E_m(x) = -\frac{1}{\beta}\ln{\{e^{-\frac{1}{2}\beta\kappa^L_m(x+x_m)^2} + e^{-\frac{1}{2}\beta\kappa^R_m(x-x_m)^2 + \Delta E}\}}
\end{equation}
where $\beta$ is the usual thermal energy; $\kappa^{L/R}$ is the curvature of the left/right well; $x$ is the particle position, $\pm x_m$ are the distances from the potential minima to the barrier; and $\Delta E$ is the free energy difference between the `folded' (left well) and `unfolded' (right well) states. The full potential the particle is subject to is given by $E_m(x)$ plus a harmonic trap of stiffness $k_s$ with a time dependent minimum location $\lambda(t)$:
\begin{equation}
    E(x, \lambda(t)) = \frac{1}{2}k_s(x-\lambda(t))^2 + E_m(x)
    \label{eq:bistableE}
\end{equation}
Following Sivak and Crooks~\cite{sivakThermodynamicGeometryMinimumdissipation2016}, we set $\kappa^L_m = \kappa^R_m = \kappa_m$, $x_m = 10 \si{\nm}$, and $\Delta E = 0$. We simulate two landscape barrier heights, $E_\textup{B} = E_m(0)-E_m(x_m) = 2.5\kT$ and $10\kT$. Setting the barrier height $E_\textup{B} = E_m(0)-E_m(x_m)$ to be 2.5$\kT$ and 10 $\kT$ fixes the curvature of the wells to be $\kappa_m = 0.2627$ and $\kappa_m = 0.8798$, respectively.

In carrying out overdamped Langevin (i.e. Brownian) dynamics simulations according to (\ref{eq:langevin}), we again follow Sivak and Crooks~\cite{sivakThermodynamicGeometryMinimumdissipation2016} and use a diffusion coefficient, spring constant, and mass representative of the dielectric beads used in optical tweezer experiments: $D = \num{0.44}\si{\D}$, $k_s = 0.4 \si{\pN\per\nm}$ and mass = \num{1e-17}\si{\gram}. Our simulations are carried out at room temperature, $\kT = \frac{1}{\beta} = 4.114 \si{\pNnm}$, and friction coefficient $\gamma = \frac{\kT}{D m}$ as per the Einstein relation.

We use an integration time step of $\delta t = 10^{-6}$ \si{\second} and equilibrate the particle for $0.001$ \si{\second} prior to beginning the protocol, which moves the trap from $x = -10$\si{\nm} to $x = 10$\si{\nm} in $t_f=10$ \si{\milli\second}.

In practice, we found that using the REINFORCE method provided better convergence than straight computation of $\langle \nabla_{\bm{\theta}} \omega \rangle$ for the 10kT landscape; this effect has been noted elsewhere~\cite{parmasPIPPSFlexibleModelBased2019}, and so we explicitly compute both terms in (\ref{eq:reinforce}) for this case.

We optimize on batches of N = 2504 trajectories with the Adam optimizer and parameters $b1=0.9$, $b2=0.999$, $eps=1\times10^{-8}$, and a learning rate that begins at $\alpha=0.1$ and decays exponentially to $\alpha=0.001$ by the end of the optimization, which we run for 1000 iterations. As an initial guess, we use a linear protocol. Our protocols are defined piecewise linearly between the first ($t=0$) and second ($t=\delta t$) points and between the penultimate ($t=t_f - \delta t$) and final ($t=t_f$) points and parametrized between $\delta t$ and $t_f-\delta t$ using a degree 13 Chebyshev polynomial; this allows us to capture any discrete jumps at the beginning and end of the protocol. 

\section*{Acknowledgements}
We thank Carl Goodrich, Ella King, and Daniel Fisher for important conversations and Grant Rotskoff for generously sharing data with us. This research was supported by the Office of Naval Research through ONR N00014-17-1-3029 and the Simons Foundation. M.C.E. also thanks Schmidt Futures in partnership with The Rhodes Trust for funding this work.

\bibliographystyle{unsrt}  
\bibliography{non_eqm_paper.bib}  %%% Remove comment to use the external .bib file (using bibtex).

\end{document}

\maketitle
\section{Supplemental text}
\subsection*{Ising MC simulations}
The best hyperparameters for optimization depend strongly on the particulars of the problem being solved. Here, we used an initial protocol guess of $B(t) = B_0(1 - t) + B_ft$, $\ln T(t) = \ln( 4t(1-t)(T_{max}-T_{min}) + T_{min}) $, where $B_0, B_f$ are the initial and final fields, respectively (-1.0 and 1.0 here) and $T_{max}, T_{min}$ are the maximal and minimal temperatures, respectively (we used 0.65 and 10). We then fit 32-degree Chebyshev polynomials to the deviation from this initial guess.

For the t=5000 optimization, we used the AD optimal curve as an initial guess. For the t=10000 optimization, we used the t=5000 AD optimal curve as an initial guess.

\begin{figure}
    \centering
    \includegraphics[width=\columnwidth]{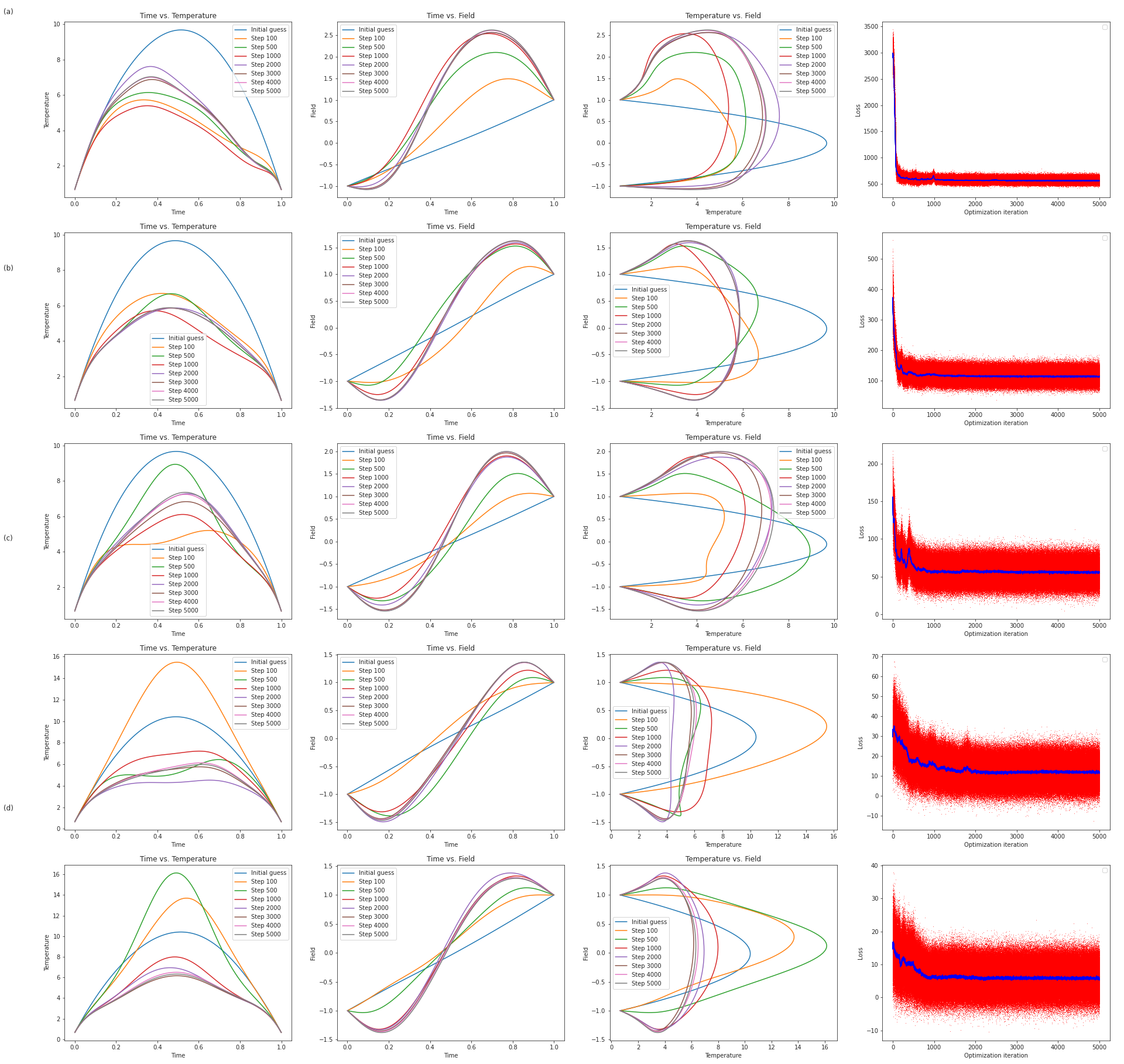}
    \caption{Convergence of the optimal protocol and entropy dissipation (loss) as a function of Adam optimization iteration for the (a) t=50, (b) t=100, (c) t=500, and (d) t=1000 simulation lengths. The initial guess curves are quadratic in temperature and linear in magnetic field, and the AD curves in the text are the result of N=5000 optimization iterations. Gradients are averaged over batches of n=256 simulations.}
    \label{fig:ising_losses_one}
\end{figure}

\begin{figure}
    \centering
    \includegraphics[width=\columnwidth]{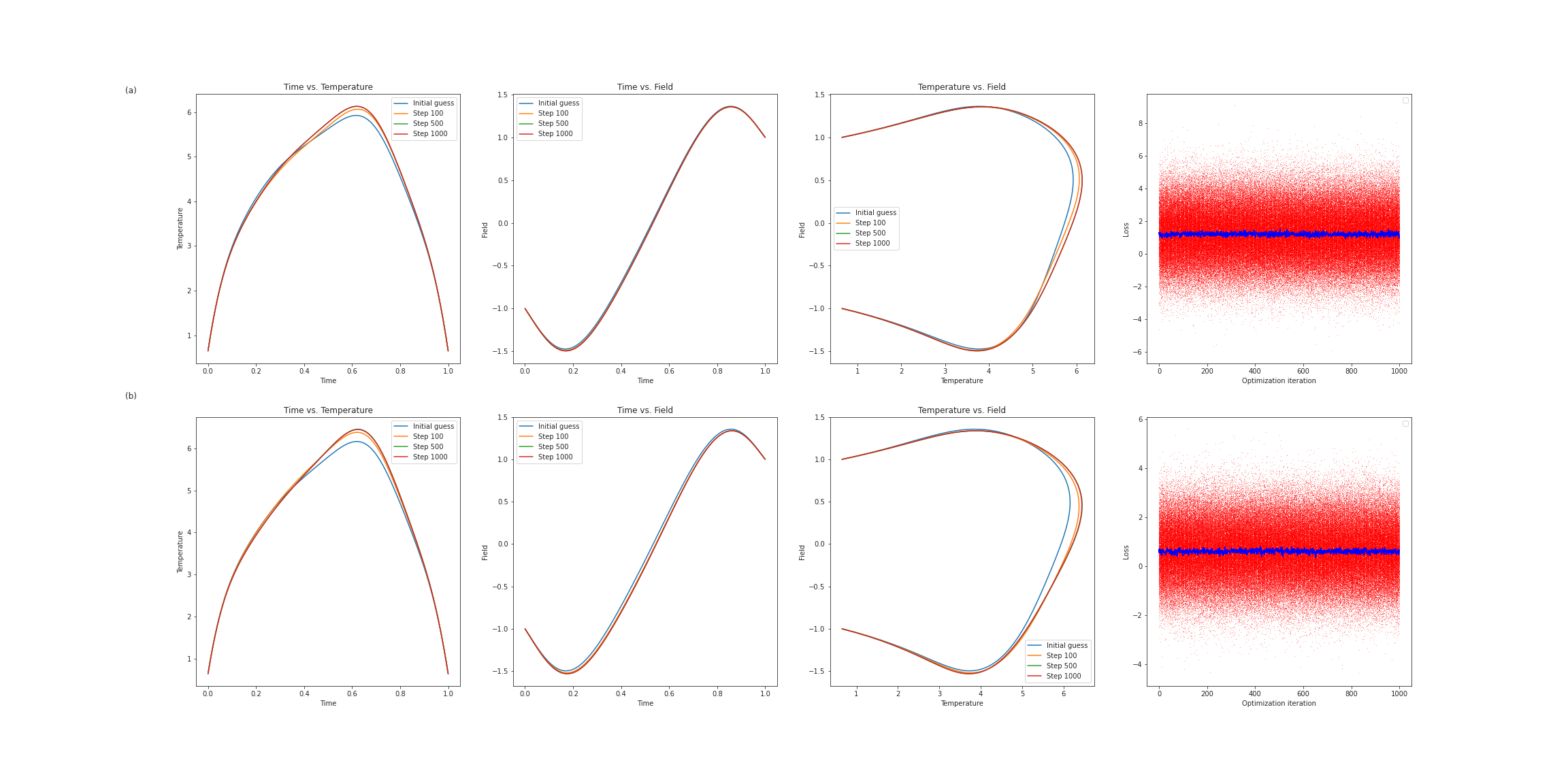}
    \caption{Convergence of the optimal protocol and entropy dissipation (loss) as a function of Adam optimization iteration for the (a) t=5000 and (b) t=10000 simulation lengths. For the t=5001 simulation, the t=1000 optimal curve is used as an initial guess and no significant decrease in loss is observed after N=1000 optimization iterations. For the t=10000 simulations, the t=5000 optimal curve is used as an initial guess, and again, no significant decrease in loss is observed after N=1000 optimization iterations. Gradients are averaged over batches of n=256 simulations. }
    \label{fig:ising_losses}
\end{figure}

The convergence of entropy production with optimization iteration is shown in Figures \ref{fig:ising_losses_one} and \ref{fig:ising_losses}.

%The full derivation for the REINFORCE gradient used is as follows: 
%Using the product rule on the integrand, the gradient of this average is 

%\begin{equation}
%\begin{split}
%    \nabla_{\bm{\theta}} \langle \omega(\bm{\theta})\rangle &= \int \mathcal{D}\bm{x}(t) (\nabla_{\bm{\theta}} P) \omega + P (\nabla_{\bm{\theta}} \omega) \\
%    &= \int \mathcal{D}\bm{x}(t) (P \nabla_{\bm{\theta}} \ln P) \omega + P (\nabla_{\bm{\theta}} \omega) \\
%    &= \langle (\nabla_{\bm{\theta}} \ln P) \omega \rangle + \langle \nabla_{\bm{\theta}} \omega \rangle
%\end{split}
%\label{eq:reinforce1}
%\end{equation}

%%TEMP:

%\begin{equation}
%    \langle \omega(\bm{\lambda})\rangle = \int \mathcal{D}\bm{x}(t) P[\bm{x}(t), \bm{\lambda}] \omega[\bm{x}(t), \bm{\lambda}]
%\end{equation}

%\begin{equation}
%\begin{split}
%    \nabla_{\bm{\lambda}} \langle \omega(\bm{\lambda})\rangle &= \int \mathcal{D}\bm{x}(t) (\nabla_{\bm{\lambda}} P) \omega + P (\nabla_{\bm{\lambda}} \omega) \\
%    &= \int \mathcal{D}\bm{x}(t) (P \nabla_{\bm{\lambda}} \ln P) \omega + P (\nabla_{\bm{\lambda}} \omega) \\
%    &= \langle (\nabla_{\bm{\lambda}} \ln P) \omega \rangle + \langle \nabla_{\bm{\lambda}} \omega \rangle
%\end{split}
%\label{eq:reinforce3}
%\end{equation}
\subsection*{Brownian dynamics}

Sivak and Crooks~\cite{sivakThermodynamicGeometryMinimumdissipation2016} consider a Brownian particle diffusing on the bistable potential 
\begin{equation}
    E_m(x) = -\frac{1}{\beta}\ln{\{e^{-\frac{1}{2}\beta\kappa^L_m(x+x_m)^2} + e^{-\frac{1}{2}\beta\kappa^R_m(x-x_m)^2 + \Delta E}\}}
\end{equation}
where $\beta$ is the usual thermal energy; $\kappa^{L/R}$ is the curvature of the left/right well; $x$ is the particle position, $\pm x_m$ are the distances from the potential minima to the barrier; and $\Delta E$ is the free energy difference between the `folded' (left well) and `unfolded' (right well) states. The full potential the particle is subject to is given by $E_m(x)$ plus a harmonic trap with a time dependent minimum location $\lambda(t)$ (simulating optical tweezers, for example) and driving barrier crossing:
\begin{equation}
    E(x, \lambda(t)) = \frac{1}{2}k_s(x-\lambda(t))^2 + E_m(x)
    \label{eq:bistableE}
\end{equation}
where $k_s$ is the stiffness of the harmonic trap. 

Within the near-equilibrium formalism of Sivak and Crooks~\cite{sivakThermodynamicMetricsOptimal2012a,sivakThermodynamicGeometryMinimumdissipation2016}, the minimum dissipation protocols are inversely proportional to the square root of the friction coefficient on the potential energy surface:
\begin{equation}
    \frac{d\lambda^{opt}}{dt} \propto \zeta^{-1/2},
    \label{eq:lambda_ODE}
\end{equation}
where the proportionality constant is fixed by the duration of the protocol. 
The friction coefficient $\zeta$ in the case of one-dimensional Brownian dynamics can be computed analytically~\cite{ZulkowskiDeweese2015}:
\begin{equation}
    \zeta(\lambda) = \frac{1}{\beta D}\int_{-\infty}^{\infty} dx \left[\frac{\left(\partial_\lambda \Pi_{eq}(x, \lambda)\right)^2}{\pi_{eq}(x,\lambda)}\right]
    \label{eq:friction}
\end{equation}
where $\pi_{eq}(x,\lambda)$ is the equilibrium probability distribution, given for the potential in Eqn. (\ref{eq:bistableE}) by~\cite{sivakThermodynamicGeometryMinimumdissipation2016}:
%\begin{multline*}
\begin{equation}
    \pi(x,\lambda) = \\ \frac{1}{Z}e^{-\frac{1}{2}\beta k_s(x-\lambda)^2} \left(e^{-\frac{1}{2}\beta \kappa_m^L(x+x_m)^2} + e^{-\beta \left[\Delta E + \frac{1}{2}\kappa_m^R(x-x_m)^2\right]}\right)
    \end{equation}
%    \end{multline*}
with $Z$ the partition function:
\begin{equation}
    Z = \sqrt{\frac{2\pi}{\beta}}\left(\frac{e^{-\frac{1}{2}\beta k_{ch}^L(\lambda+x_m)^2}}{\sqrt{k_s + \kappa_m^L}} + \frac{e^{-\beta \left[\Delta E + \frac{1}{2}k_{ch}^R(\lambda-x_m)^2 \right]}}{\sqrt{k_s + \kappa_m^R}} \right)
\end{equation}
and $k_{ch}^{L/R}$ a characteristic spring constant given by: 
\begin{equation}
    k_{ch}^{L/R} \equiv \frac{k_s\kappa_m^{L/R}}{k_s + \kappa_m^{L/R}};
\end{equation}
and with $\Pi_{eq}(x, \lambda)$ the cumulative distribution function~\cite{ZulkowskiDeweese2015}:
\begin{equation}
    \Pi_{eq}(x, \lambda) = \int_{-\infty}^xdx' \pi(x',\lambda)
\end{equation}

To reproduce the optimal protocols calculated by Sivak and Crooks\cite{sivakThermodynamicGeometryMinimumdissipation2016}, we computed $\Pi_{eq}(x, \lambda)$ and its derivative with respect to trap position, $\partial_\lambda \Pi_{eq}(x, \lambda)$, in Mathematica, then numerically integrated Eqn. (\ref{eq:friction}) with the trapezoid integrator in Numpy.

Once $\zeta(\lambda)$ is known, the differential equation (\ref{eq:lambda_ODE}) can be solved using trapezoidal integration to yield the optimal protocol $\lambda(t)$. 

\section{Supplemental figures}
%%% Each figure should be on its own page
\newpage
\newpage

\newpage
\begin{figure}
    \centering
    \includegraphics[width=\columnwidth]{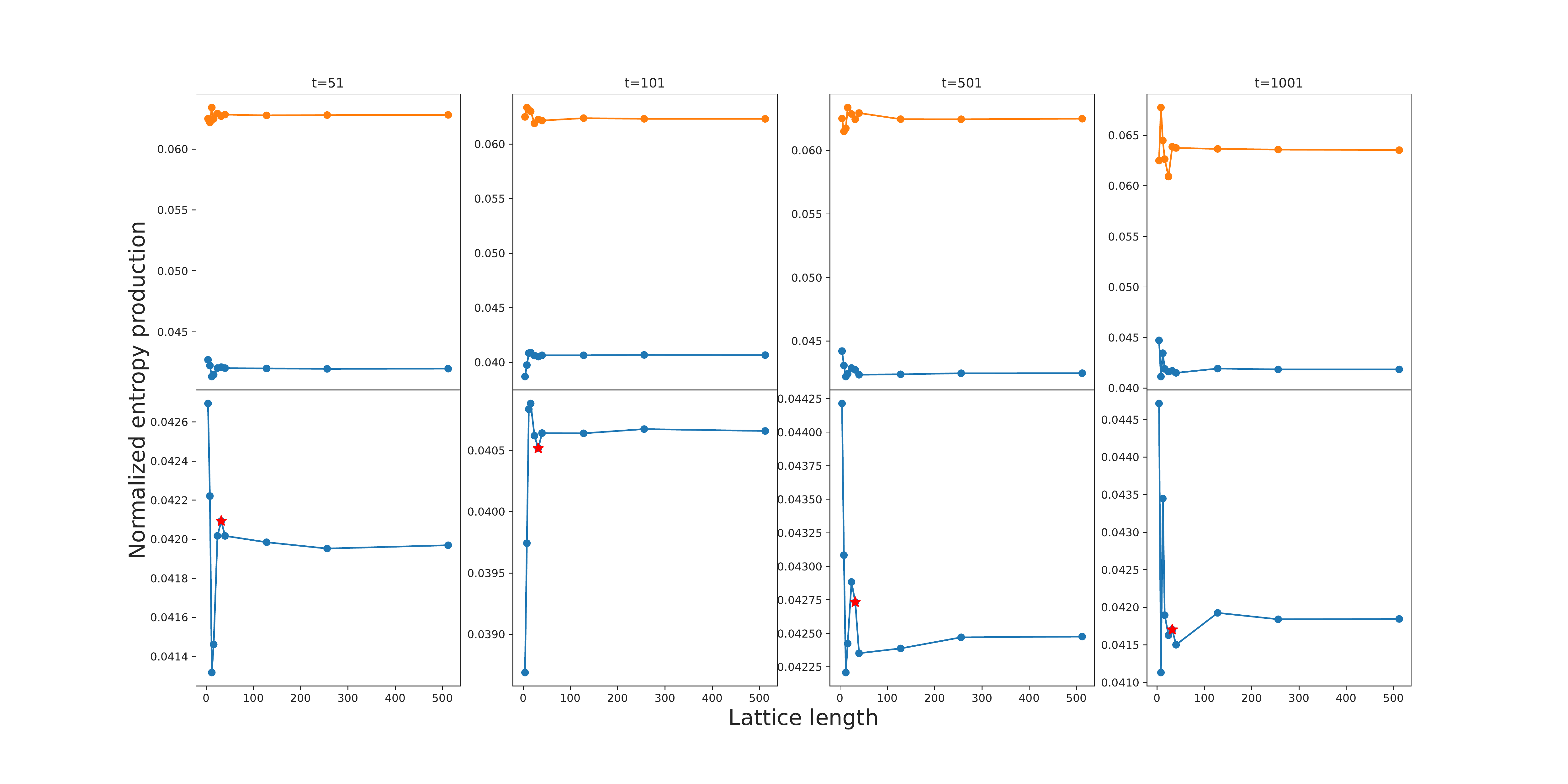}
    \caption{Normalized entropy production as a function of lattice size (4x4, 8x8, 12x12, 16x16, 24x24, 32x32, 40x40, 128x128, 256x256, and 512x512 shown). Here, the optimal protocol derived using automatic differentiation on a 32x32 lattice is used to run forward simulations on smaller and larger lattices to test whether optimizing on a 32x32 yields a good approximation of the infinite-lattice optimal curve. Forward simulations are also run using the near-equilibrium (NE) theoretical optimal curve of Sivak and Crooks~\cite{rotskoffOptimalControlNonequilibrium2015}. For each simulation duration plotted (t=50, 100, 500, 1000), the entropy production is normalized I. for lattice size, by dividing by lattice area, and II. for the fact that entropy dissipation decreases with for longer (i.e. closer-to-equilibrium) simulations, by dividing by entropy production for a t=50 simulation on a 4x4 lattice using the near-equilibrium (NE) theoretical optimal protocol of Sivak and Crooks~\cite{rotskoffOptimalControlNonequilibrium2015}. Top panel: A comparison between the entropy produced with the AD-derived optimal protocol (blue) and the near-equilibrium (NE) theoretical protocol (orange) for each simulation length, showing that the AD-derived protocols outperform near-equilibrium theory in the infinite-lattice limit; Bottom panel: zoom-in on the results of the AD-derived protocols, showing that the normalized entropy production has converged to the infinite-lattice limit. The 32x32 result, used in the main text, is shown as a red star, and agrees to within 1\% of the infinite-lattice limit. Entropies are averages over N=2560 simulations.
     }
    \label{fig:ising_convergence}
\end{figure}
\newpage
\begin{figure}
    \centering
    \includegraphics[width=\columnwidth]{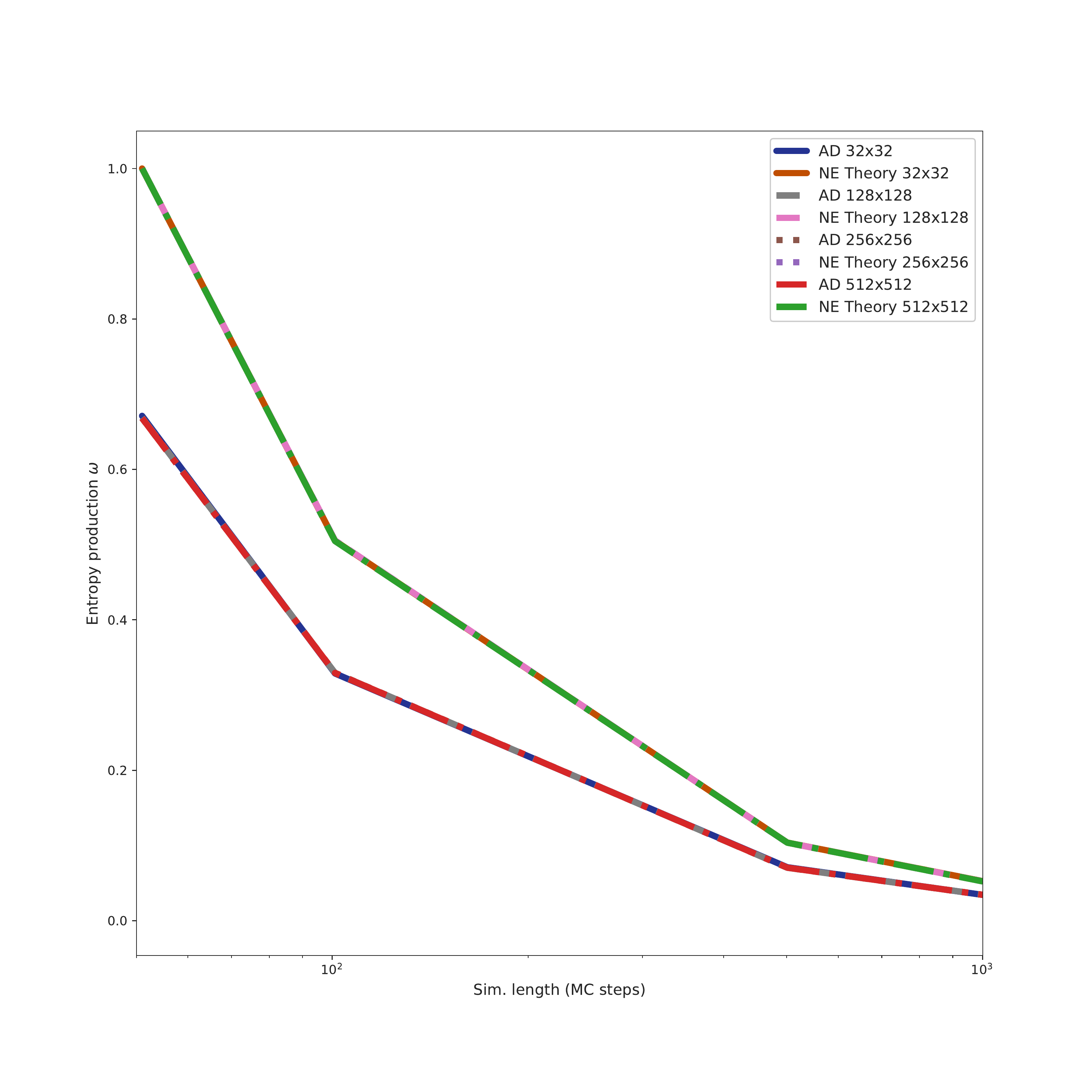}
    \caption{Normalized entropy production (all entropies are divided by the near-equilibrium (NE) theoretical protocol entropy dissipation for t=50) for four different lattice sizes: 32x32 (featured in the main text), 128x128, 256x256, and 512x512, for both the near-equilibrium (NE) theoretical optimal protocol of Sivak and Crooks~\cite{rotskoffOptimalControlNonequilibrium2015} and for the optimal protocol derived using automatic differentiation (AD) on 32x32 lattice simulations. Results for larger lattices collapse onto the 32x32 results, demonstrating that performing AD optimization on a smaller lattice provides a good estimation of the optimal protocol in the infinite lattice limit.}
    \label{fig:ising_lattice_sizes}
\end{figure}
\newpage
\begin{figure}
    \centering
    \includegraphics[width=\columnwidth]{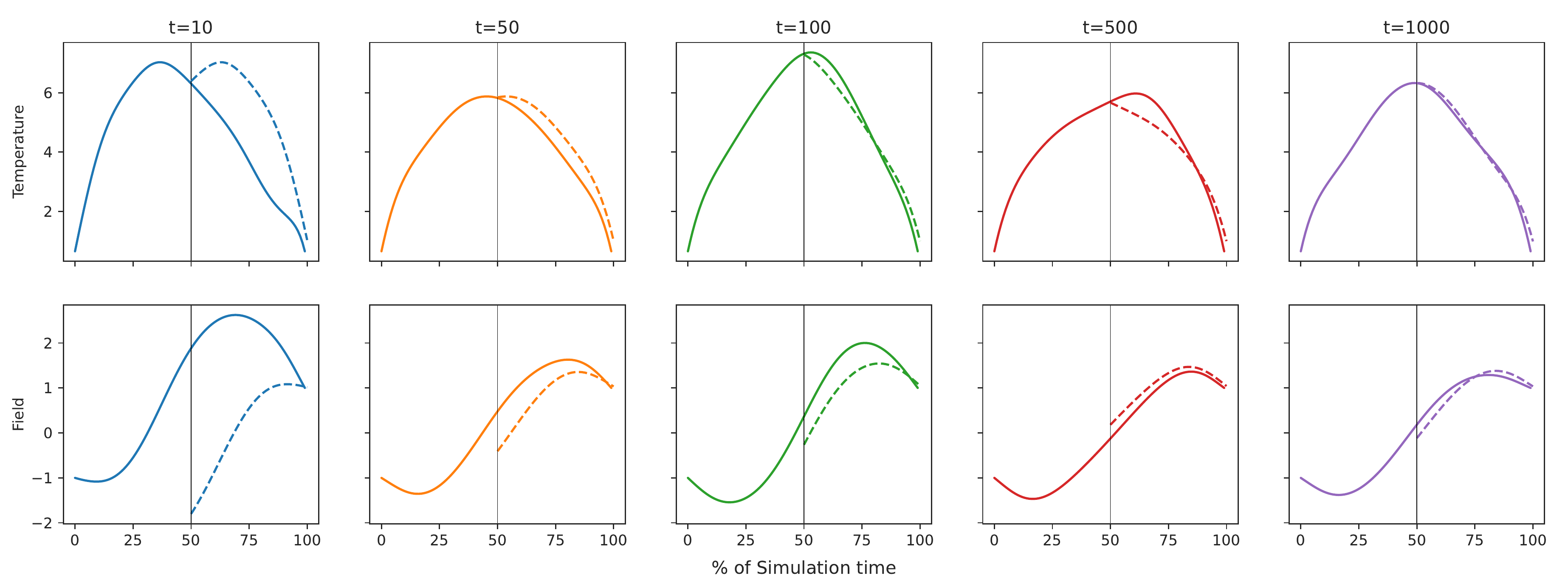}
    \caption{Symmetry breaking in the optimal protocols derived with automatic differentiation. Protocols are the result of automatic differentiation based optimization on a 32x32 lattice, for the four simulation lengths shown. By contrast, the near-equilibrium (NE) theoretical optimal protocol is necessarily time symmetric~\cite{rotskoffOptimalControlNonequilibrium2015}. Dotted lines are a guide for the eye and indicate what a symmetric reflection of the first half of each protocol would look like. In general, protocol symmetry increases as equilibrium is approached, as expected.}  
    \label{fig:ising_symmetry}
\end{figure}
\newpage
\begin{figure}
    \centering
    \includegraphics[width=\columnwidth]{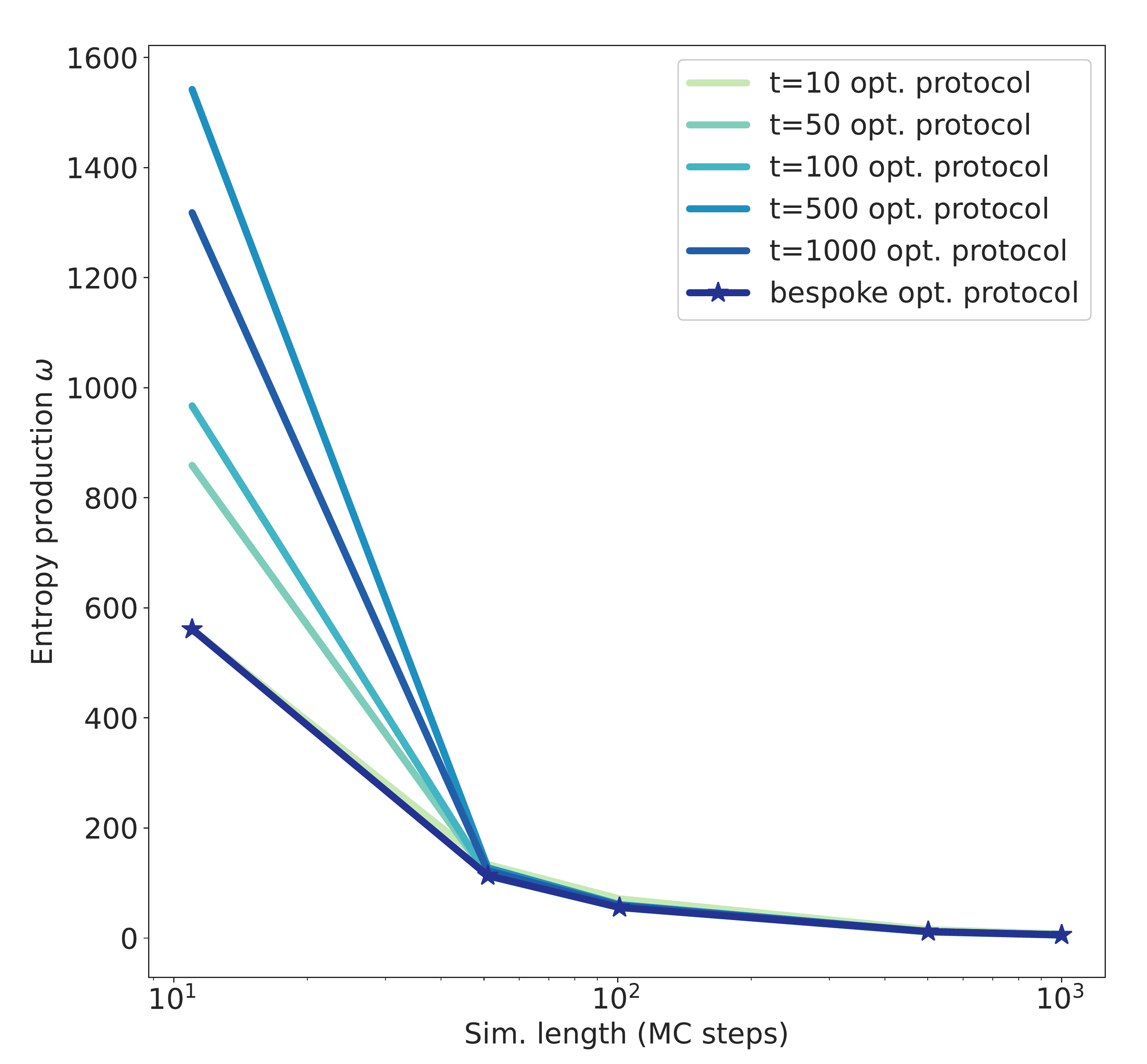}
    \caption{Performance of optimal protocols derived for a specific simulation length (t=10, 50, 100, 500, 1000) on simulations of other lengths. Specifically, the optimal protocol found using automatic differentiation for t=10 magnetization reversal simulations was run on batches of n=2560 forward simulations of lengths t=50, 100, 500, and 1000, and similarly for the other simulation lengths. The 'bespoke' curve indicates the results of using the AD-derived protocol for the specific simulation length. Notably, far-from-equilibrium optimal protocols perform comparably to the near-equilibrium protocols in the near-equilibrium regime (the reverse is not true). Standard errors of the mean are less than line widths.}
    \label{fig:ising_performance}
\end{figure}
\newpage
\begin{figure}
    \centering
    \includegraphics[width=\columnwidth]{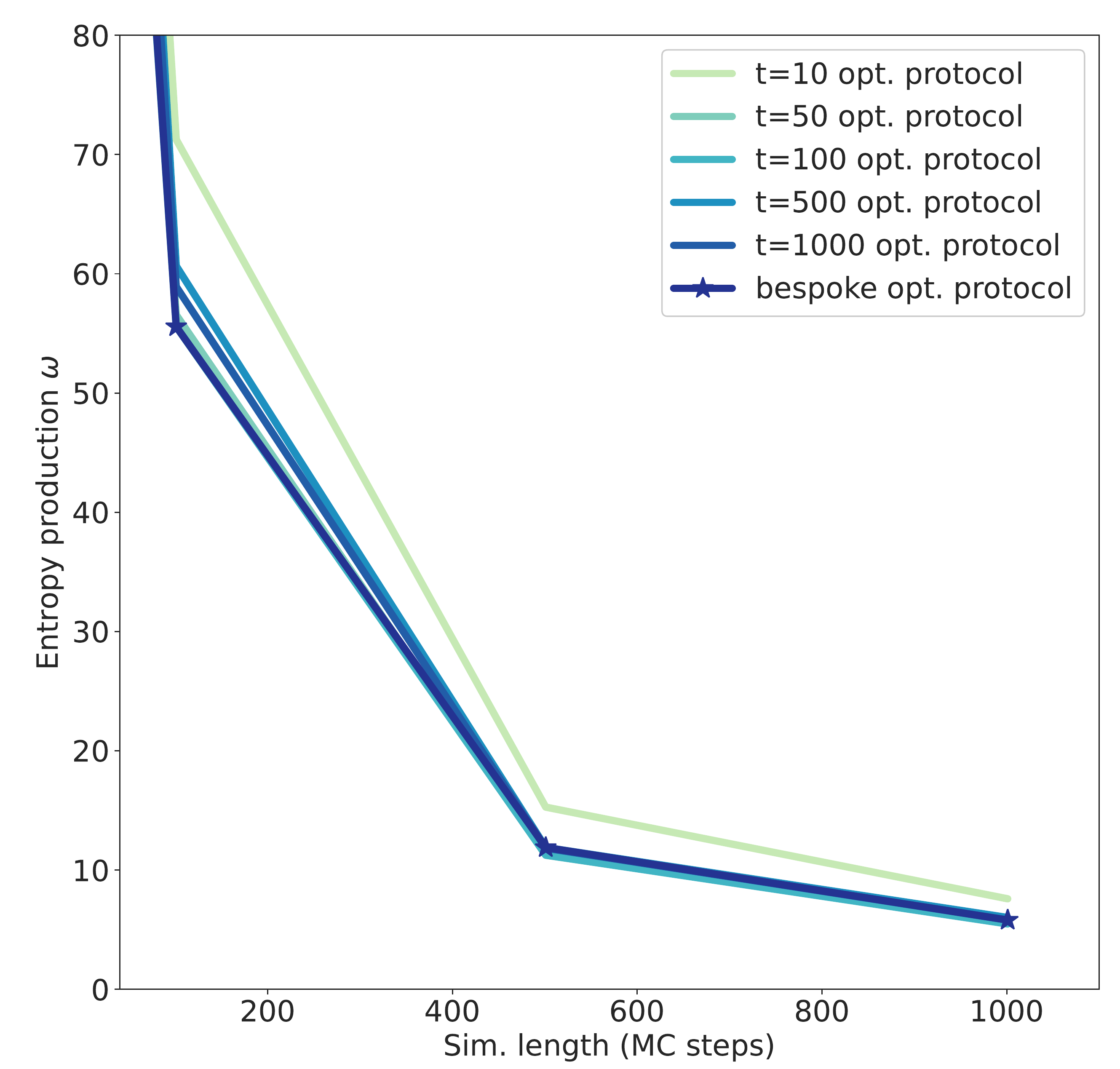}
    \caption{Same as above, but zoomed in on the near-equilibrium region. Performance of optimal protocols derived for a specific simulation length (t=10, 50, 100, 500, 1000) on simulations of other lengths. Specifically, the optimal protocol found using automatic differentiation for t=10 magnetization reversal simulations was run on forward simulations of lengths t=50, 100, 500, and 1000, and similarly for the other simulation lengths. The 'bespoke' curve indicates the results of using the AD-derived protocol for the specific simulation length. Notably, far-from-equilibrium optimal protocols perform comparably to the near-equilibrium protocols in the near-equilibrium regime (the reverse is not true). Standard errors of the mean are less than line widths.}
    \label{fig:ising_performance_zoom}
\end{figure}
\newpage

\begin{figure}
    \centering
    \includegraphics[width=\columnwidth]{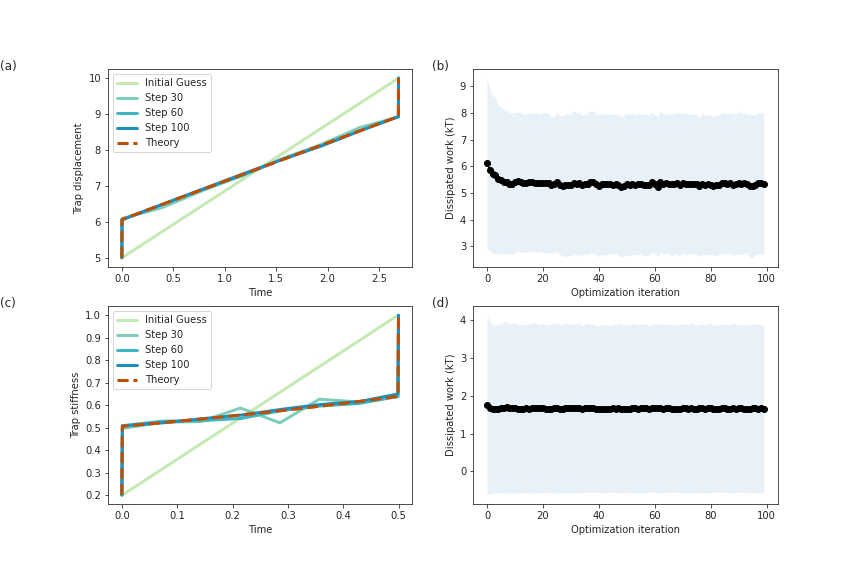}
    \caption{Convergence of the automatic differentiation-derived optimal protocols to the exact theoretical optimal protocols of Schmiedl and Seifert~\cite{schmiedlOptimalFiniteTimeProcesses2007} for a Brownian particle in (a) a moving harmonic trap and (c) a harmonic trap with time-varying stiffness, as shown in the main text. Loss (dissipated work) convergence is also shown for (b) the moving trap and (c) the trap varying in stiffness.}
    \label{fig:brownian_trap}
\end{figure}
\newpage
\begin{figure}
    \centering
    \includegraphics[width=\columnwidth]{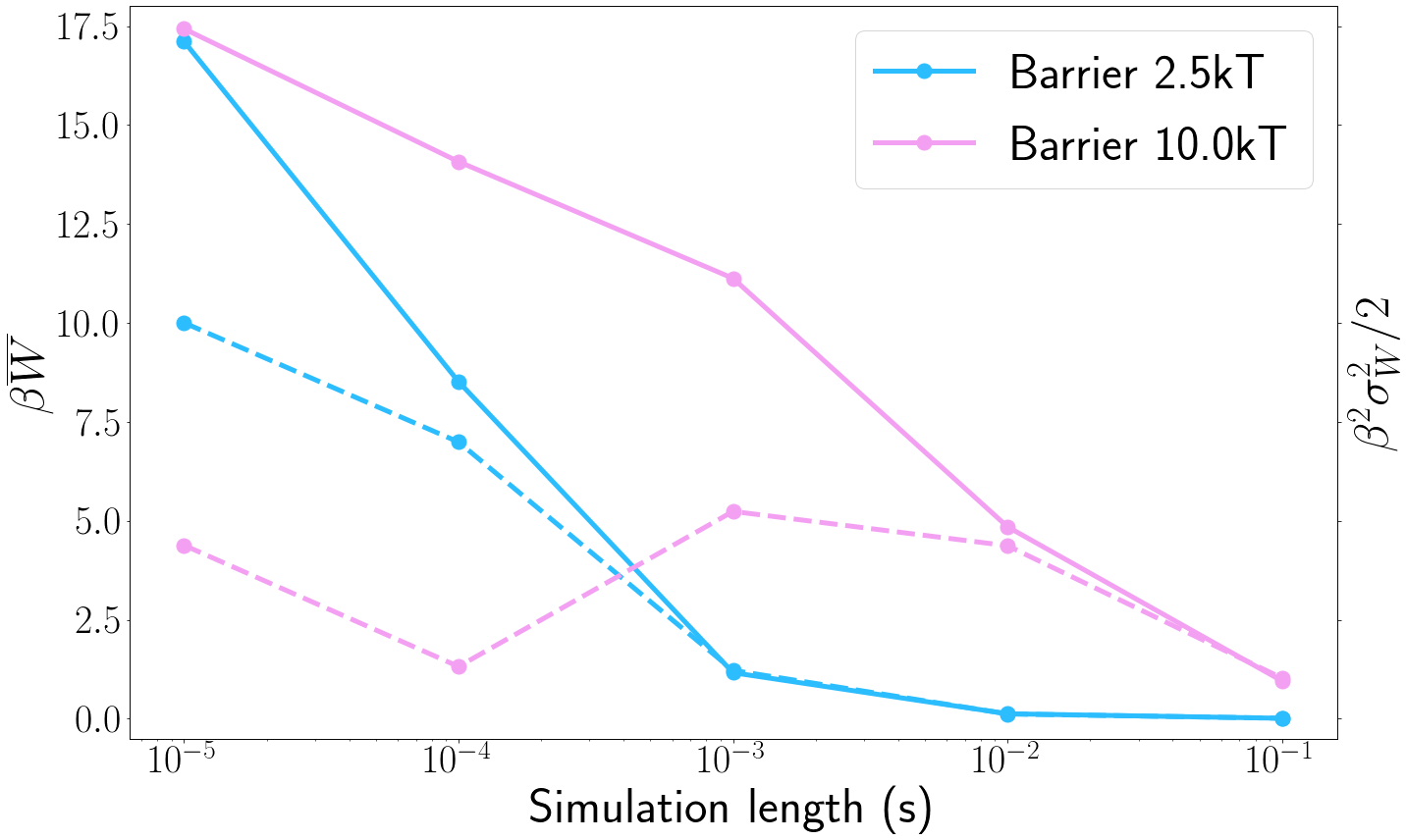}
    \caption{Average \emph{cyclical} dissipated work (\textbf{solid lines}, left ordinate) and half the variance in cyclical dissipated work (\textbf{dotted lines}, right ordinate) for barrier crossing simulations of various total lengths. For these cyclical simulations, a Brownian particle is dragged across its landscape and then \emph{back} to its original position. The simulation times in the abscissa are half of the total cycle time, and longer simulations are closer to equilibrium. At or near equilibrium, the average cycle work is expected to equal half the variance~\cite{hermansSimpleAnalysisNoise1991}. Thus, for simulations perfectly at equilibrium, we expect the dotted and solid lines to overlap. Here, the simulations on a 2.5\si{\kT} barrier satisfy the $\beta\overline{W} = \beta^2\sigma_W^2/2$ condition for simulations longer than 1\si{\milli\second}, while the 10\si{\kT} barrier simulations begin to deviate from this condition already at 10\si{\milli\second} simulation lengths. Here, a a batch of 5000 simulations is run at each simulation end time, and a linear trap protocol -- not an optimal protocol -- is used. All other parameters used are as described in Materials \& Methods in the main text. The main text features simulations of length t=10\si{\milli\second}, where the 2.5\si{\kT} barrier landscape dynamics are near-equilibrium and the 10\si{\kT} barrier landscape dynamics have begun to deviate from equilibrium.}
    \label{fig:cycle_work}
\end{figure}
\newpage
\begin{figure}
    \centering
    \includegraphics[width=\columnwidth]{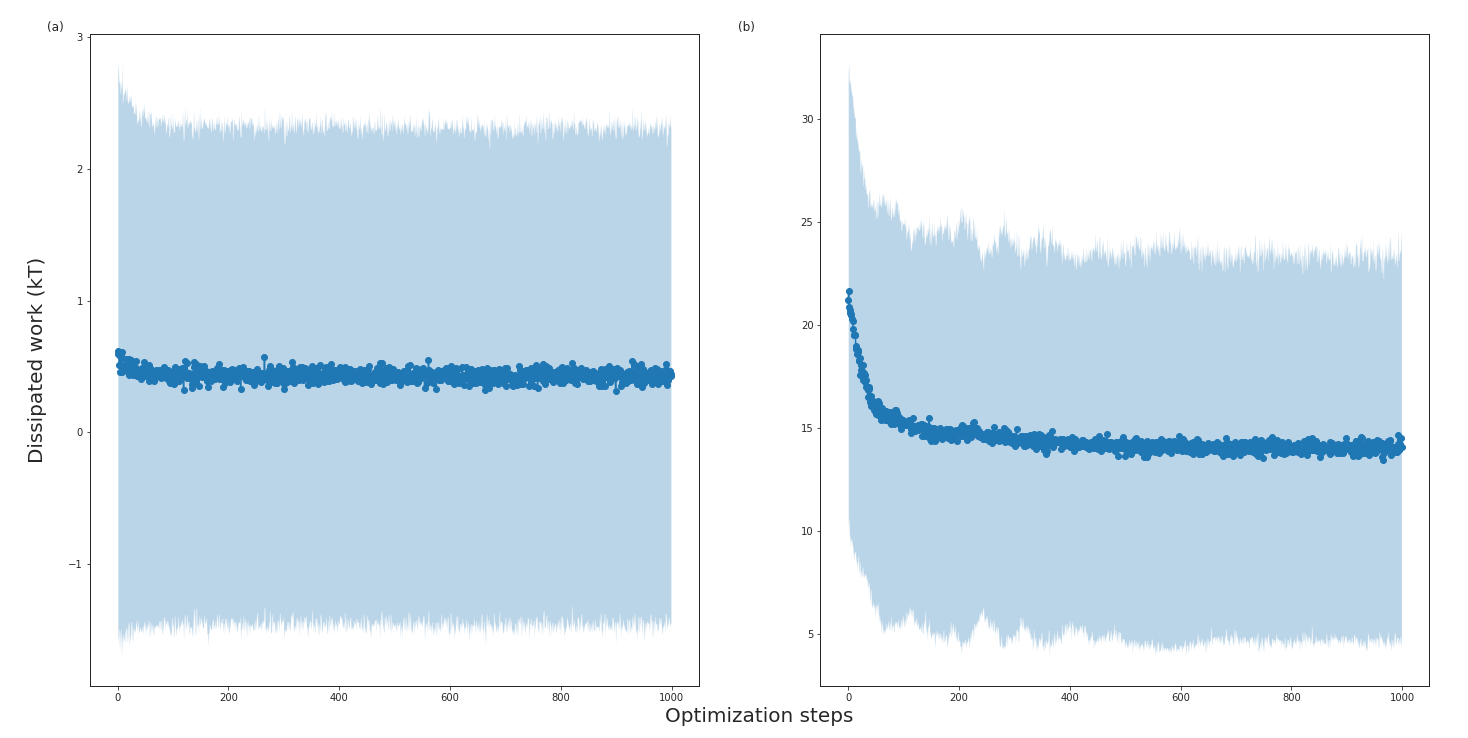}
    \caption{Loss versus Adam optimization iteration for the (a) 2.5\si{\kT} barrier and (b) 10\si{\kT} barrier optimal protocols shown in the main text.}
    \label{fig:bar_losses}
\end{figure}

\bibliographystyle{unsrt}  
\bibliography{non_eqm_paper.bib}  %%% Remove comment to use the external .bib file (using bibtex).